%% file: draft_main.tex
\UseRawInputEncoding 

\documentclass[%
 reprint,
 amsmath,amssymb,
 aps,
pra,
]{revtex4-2}

\usepackage{graphicx}
\usepackage{dcolumn}
\usepackage{bm}
\usepackage{stackengine}

\input{header_macros.tex} 

\begin{document}

\preprint{APS/123-QED}

\title{Strong dispersive coupling between a mechanical resonator and a fluxonium superconducting qubit}

\author{Nathan R. A. Lee}
 \email{nlee92@stanford.edu} 

\author{Yudan Guo}

\author{Agnetta Y. Cleland}

\author{E. Alex Wollack}

\author{Rachel G. Gruenke}
\author{Takuma Makihara}
\author{Zhaoyou Wang}

\author{Taha Rajabzadeh}

\author{Wentao Jiang}

\author{Felix M. Mayor}

\author{Patricio Arrangoiz-Arriola}

\author{Christopher J. Sarabalis}

\author{Amir H. Safavi-Naeini}
 \email{safavi@stanford.edu} 

\affiliation{%
 Department of Applied Physics and Ginzton Laboratory, Stanford University\\
 348 Via Pueblo Mall, Stanford, California 94305, USA
}

\date{\today}

\begin{abstract}
We demonstrate strong dispersive coupling between a fluxonium superconducting qubit and a 690 megahertz mechanical oscillator, extending the reach of circuit quantum acousto-dynamics (cQAD) experiments into a new range of frequencies. We have engineered a qubit-phonon coupling rate of $g\approx2\pi\times14~\text{MHz}$, and achieved a dispersive interaction that exceeds the decoherence rates of both systems while the qubit and mechanics are highly nonresonant ($\Delta/g\gtrsim10$). Leveraging this strong coupling, we perform phonon number-resolved measurements of the mechanical resonator and investigate its dissipation and dephasing properties. Our results demonstrate the potential for fluxonium-based hybrid quantum systems, and a path for developing new quantum sensing and information processing schemes with phonons at frequencies below 700 MHz to significantly expand the toolbox of cQAD.
\end{abstract}

\maketitle

\section{\label{sec:intro} Introduction}
Mechanical resonators are promising candidates for hardware-efficient quantum memory \cite{Hann2019, Pechal2019, Chamberland2022} and novel types of quantum sensors, as they offer a smaller spatial footprint and couple to degrees of freedom such as mass and force. In this work, we achieve strong dispersive coupling between a fluxonium superconducting qubit and a sub-gigahertz mechanical oscillator -- an essential step toward developing quantum sensing and networking components operating at lower frequencies. Achieving significant coupling between qubits and mechanical oscillators poses a challenge within the established paradigm pursued in most recent circuit quantum acoustodynamics (cQAD) efforts, where piezoelectricity mediates resonant coupling between a weakly anharmonic transmon qubit and a mechanical oscillator~\cite{Manenti2017, Noguchi2017, Noguchi2020, Moores2018, Kervinen2018, Sletten2019, Oconnell2010, Chu2017, Chu2018, VonLupke2022, Bild2022, Arrangoiz-Arriola2016, Arrangoiz-Arriola2018, Arrangoiz-Arriola2019, Satzinger2018, Wollack2021, Wollack2022, Cleland2023}. Most of these demonstrations operate in the 2 to 8 gigahertz frequency range. We are motivated to work with lower-frequency resonators, which are expected to exhibit longer coherence times and provide new frequency windows for operating quantum devices. Seeking to operate at lower mechanical frequencies effectively compels us to move to a different type of qubit~\cite{Lecocq2015b, Viennot2018, Ma2021} to avoid rapidly diminishing coupling rates -- particularly for nanomechanical oscillators that have minimal gate capacitance.  Our work shows that a substantial qubit-mechanics coupling rate can be achieved at lower frequency by using a fluxonium qubit. We show that the resulting large dispersive interaction rates, exceeding the decoherence rates of both systems, enable phonon number-resolved measurements of mechanical resonators through the fluxonium and time-dependent coherence measurements of the oscillator. Remarkably, we are able to achieve large dispersive cooperativities despite working at a large detuning ($\Delta/g \geq 10$) as compared to previous cQAD demonstrations. Larger detunings allow our mechanical resonator to be more effectively isolated from the qubit, and vice-versa, an aspect that will become more important for longer-lived resonances.

\begin{figure}[t]
\includegraphics{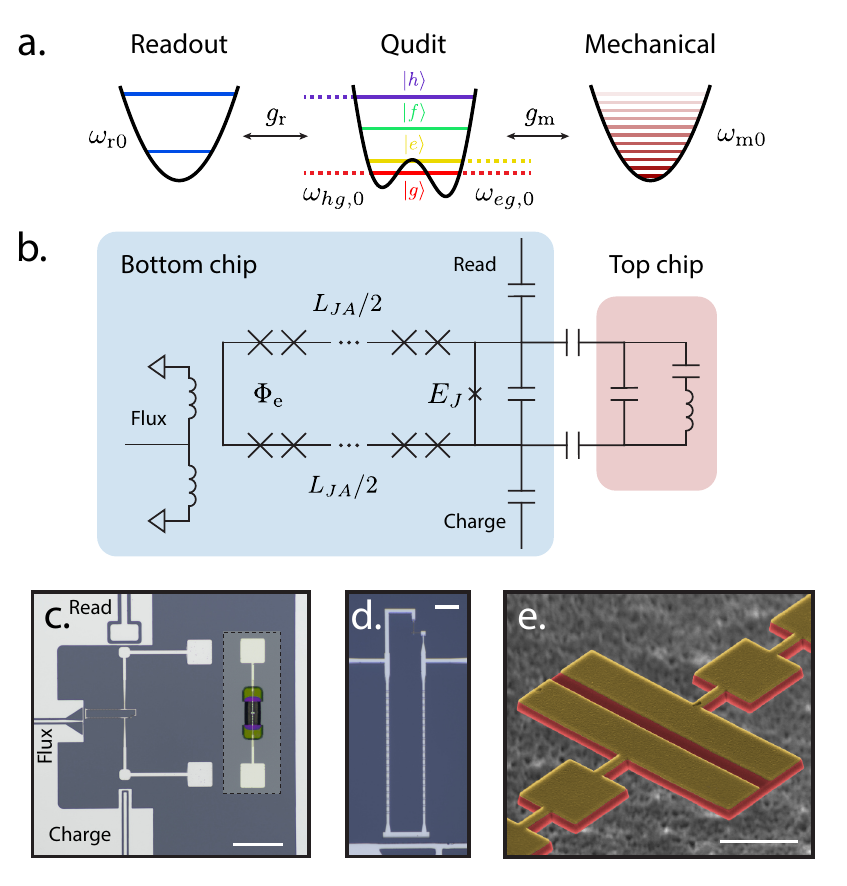}
\caption{\label{fig:f1_device_concept} \textit{Description of device}. (a) Schematic of the energy levels for different systems in this work. A strongly-nonlinear Josephson oscillator (\textit{Qudit}) is coupled to a sub-GHz mechanical resonator and a few-GHz readout resonator, with respective coupling rates $g_\text{m,r}$ between the qudit charge operator $\nqop$ and the (mechanical, readout) charge quadratures. Qudit transitions dominating these respective interactions are labeled. (b) Circuit schematic of the flip-chip device. The qudit is patterned on the bottom chip (blue) and coupled to the mechanical mode on the top chip (maroon) through two vacuum-gap capacitors. The target mechanical mode is represented as a Butterworth-van Dyke equivalent circuit, omitting additional series-$LC$ branches describing parasitic modes of the real device. (c) Optical micrograph of the qudit and control lines, with inset showing coupling pads leading to the mechanical resonator on the top chip. (d) Optical micrograph of the Josephson junction loop providing the qudit nonlinearity. (e) False-color scanning electron micrograph of a representative mechanical resonator. The experimental resonator was not imaged to minimize handling risks discussed in Appendix \ref{app:fabrication}. Scale bars for (c, d, e) respectively represent (50, 10, 2) $\mu m$.}
\end{figure}

In this work we demonstrate strong dispersive coupling between a lithium niobate (LN) phononic crystal resonator at $\wmech/\tpi \approx 690~\text{MHz}$ and a superconducting qubit, allowing us to resolve individual phonon levels. The device is composed of a fluxonium circuit capacitively coupled to an on-chip readout resonator~\cite{Manucharyan2009,Zhang2021} and heterogeneously integrated~\cite{Satzinger2018,Wollack2022} with a nanomechanical phononic crystal cavity~\cite{Arrangoiz-Arriola2018,Arrangoiz-Arriola2019}. The effective electrical circuit as well as microscope images of the key components at different scales are depicted in Fig. \ref{fig:f1_device_concept}. The mechanical frequency is approximately a factor of $3$ smaller than the similar resonators in Ref. \cite{Wollack2022}. In this frequency range, and at a resonator temperature of $\Teff \sim 30~\text{mK}$, thermal excitations are significantly more common ($\hbar \wmech / \kB T \sim 1$).
To address this lower frequency we use a ``light'' fluxonium qubit \cite{Manucharyan2009, Cottet2021} that preserves the insensitivity to charge noise and GHz-frequency readout associated with transmons \cite{Koch2007}, while also realizing a large qubit-phonon coupling rate $g/\tpi \approx 13.5~\text{MHz}$. The qubit and mechanics are fabricated on separate chips and coupled capacitively across a vacuum gap. We primarily operate the qubit in the strong dispersive regime ($\wegzero = \wmechzero + \Delta$ where $|\Delta|/g \gg 1$). The dynamics are then described approximately by an effective Hamiltonian \cite{Blais2004, Schuster2007},
\begin{equation} \label{eq:intro_Heff_disp}
    \Heff = -\half \weg \sz + \wmech \bdagb -\chim \sz \bdagb \, .
\end{equation}
Eqn. \ref{eq:intro_Heff_disp} suggests that we can perform quantum non-demolition (QND) measurements of phonon population by probing the qubit and that we can perform QND measurements of qubit population by probing the mechanics. When the shift-per-excitation $2\chim$ exceeds the linewidth of the qubit, individual phonon states become resolvable in the qubit excitation spectrum \cite{Gambetta2006} as suggested by eqn. \ref{eq:intro_Heff_disp}. In our system, $\hbar \omega / \kB T \sim 1$, and the equilibrium thermal state contains excitations above the ground state for both mechanics and qubit. Therefore we anticipate an excitation spectrum that follows the thermal distribution of the system. High fidelity gates generally require starting from a pure state, e.g., by cooling the system to its ground state \cite{Teufel2011,Chan2011} or otherwise stabilizing in a low-entropy state \cite{Divincenzo2000, Zhang2021}.
In this work we demonstrate mechanics-fluxonium coupling in the dispersive regime, and observe the thermal excitation spectrum of the system. We also perform partially-coherent operations on the initial thermal state to demonstrate feasibility of phonon-number measurement and single-phonon state preparation. With modest improvements in qubit frequency stability, discussed in Appendix \ref{app:to_improve}, we anticipate high-fidelty state preparation, and single-phonon control in this platform.

We organize our work in two parts. In Sec. \ref{sec:strongCoupling} we focus on characterizing the coupling and measuring the level structure in the dispersive regime. We first observe strong resonant coupling by tuning the qubit through the mechanics, performing two-tone spectroscopy and measuring the minimum splitting of the avoided crossing. We then detune the qubit to $\Dcoh / \geg\approx 9$ and observe phonon-number resolved transitions following coherent excitation of the mechanics \cite{Arrangoiz-Arriola2019, VonLupke2022}. In Sec. \ref{sec:phononCoherence} we focus on implementing partially coherent gates between the phonon and qubit. We tune the qubit to $\Dswap / \geg \approx 11$ and modulate the qubit frequency to swap single-photon-like states from the qubit into the mechanics. Using swap-like operations by frequency modulation, we measure energy decay ($\ToneM$) and phase decay ($\TtwoM$) of the mechanics~\cite{Wollack2022}.

\section{\label{sec:strongCoupling} Strong coupling below $1~\text{GHz}$}
\subsection{\label{subsec:resonantCoupling} Resonant coupling}
\begin{figure}[t]
\includegraphics{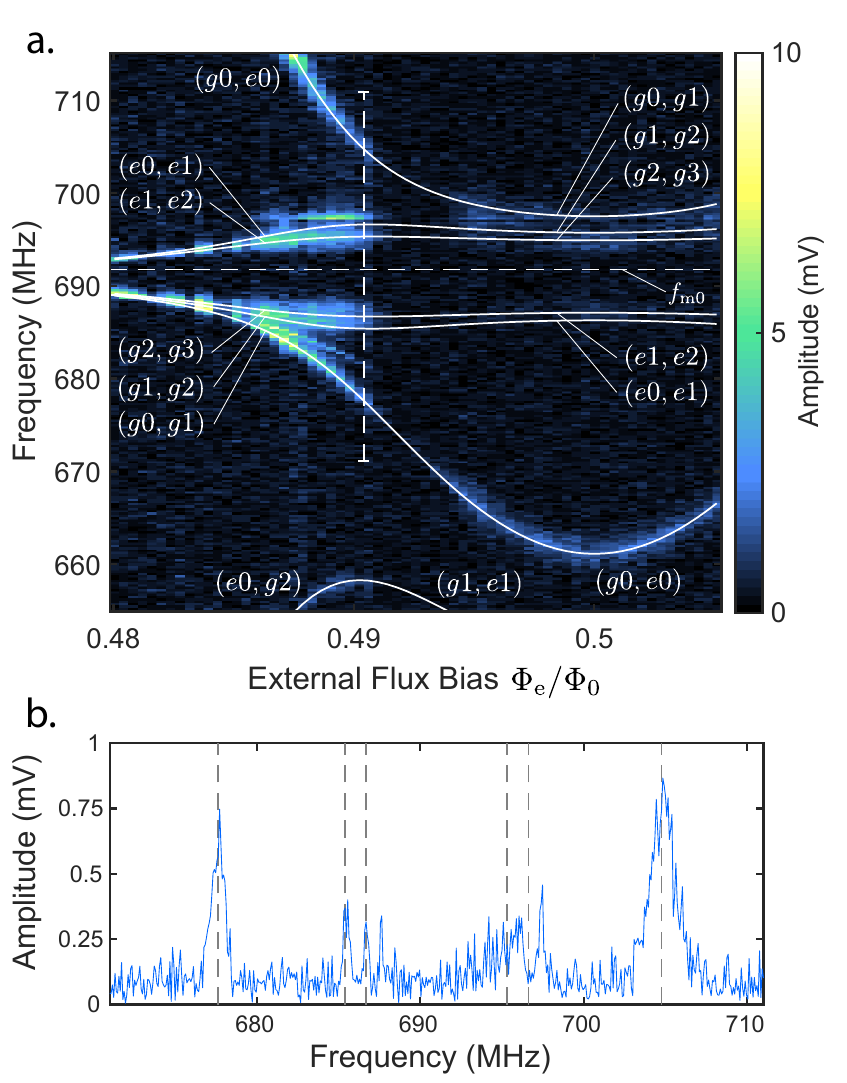}
\caption{\label{fig:f2_resonant_coupling} \textit{Spectroscopy of qubit-mechanics coupling}. (a) Qubit spectrum as a function of applied magnetic flux $\Phie$ in units of the magnetic flux quantum $\Phio$. Solid curves denote first-order transition frequencies predicted by diagonalizing the coupled qudit-mechanical Hamiltonian; the procedure for determining model parameters is detailed in Appendix \ref{app:tuning_fit}. Transitions are labeled using the \textit{undressed} basis states $\lcb|\text{qubit}\rangle \otimes |\text{mechanics}\rangle \rcb$ with greatest overlap to the eigenstates involved in each transition. These labels change when passing through the avoided crossing and are not intended as quantitative descriptions of the eigenstates, as the spectroscopy window covers a region of strong hybridization. (b) Finer spectrum taken along the vertical dashed line in (a), at the approximate center of the avoided crossing. The qubit excitation amplitude is reduced by a factor of $5$, and the unlabeled peak seen near $681.5~\text{MHz}$ in (a) no longer appears. We attribute this peak to a second-order transition \cite{Chu2017}. Vertical dashed lines denote the transition frequencies predicted in (a) and align well to the lower peaks; less well to the upper peaks.}
\end{figure}

The qudit arises from a fluxonium superconducting circuit~\cite{Manucharyan2009, Earnest2018, Lin2018, Nguyen2019, Cottet2021, Zhang2021, Somoroff2021} with Hamiltonian given by
\begin{equation} \label{eq:Hq_flxm}
    \Hq = 4 \EC \, \nqop^2 - \EJ \cos \lp \phiqop \rp + \half \EL \lp \phiqop + \phie \rp^2 ,
\end{equation}
where $\phie = \tpi \Phie / \Phio$ is the external flux bias in units of reduced flux quantum. The qudit-mechanics coupling is described by a linear piezoelectric interaction \cite{Pechal2019},
\begin{equation} \label{eq:Hint_piezo}
    \Hint = -i \hbar \gm \nqop \lp \bhat - \bdag \rp.
\end{equation}
We are primarily interested in the coupling to the qubit $(g,e)$ transition, which occurs with a rate $\geg \equiv \gm |_\text{q} \bra{g} \nqop \ket{e} _\text{q}|$. By varying the DC current flowing in the flux line, we tune the qubit frequency $\weg$ through the anticipated mechanical frequency $\wmechzero$ and observe an avoided crossing of width $2 \geg/\tpi \approx 27.1~\text{MHz}$ near $\wmechzero/\tpi \approx 692~\text{MHz}$, shown in Fig. \ref{fig:f2_resonant_coupling}a. The normalized coupling $\geg / \wmechzero \approx 1.9\%$ is on the same order as in strongly coupled superconducting-only systems \cite{Schuster2007}.

The experimental spectra we observe display features that are typically absent for transmon-mechanical avoided crossings measured with GHz-frequency mechanical resonators~\cite{Chu2018,Arrangoiz-Arriola2019}. We observe additional peaks between the outer branches of the avoided crossing. We interpret these peaks as representing transitions between levels above the ground state $\ket{g0}$, visible in spectroscopy as they are thermally excited \cite{Gely2021, Zeng2021}. To verify this interpretation, we diagonalize the original Hamiltonian \cite{Smith2016, Rajabzadeh2022} $\Hhat = \hbar\wmechzero \, \bdagb + \Hq + \Hint$ exactly and fit the energy differences between the eigenvalues to the observed peak frequencies. Except for a flux-independent feature at $697~\text{MHz}$, which we attribute to a weakly coupled parasitic mechanical resonance, the observed transition frequencies agree with the model. 
The spectrum shown in Fig. \ref{fig:f2_resonant_coupling}a is power broadened, preventing us from resolving the individual transitions. We reduce the excitation power and perform a narrower band sweep at a fixed flux of $0.49\Phio$, and observe resolved peaks near $685~\text{MHz}$ on the lower-frequency side of the window agreeing with the theoretical model (Fig. \ref{fig:f2_resonant_coupling}b). The corresponding peaks on the higher-frequency side do not agree quantitatively with the model, likely due to coupling to the parasitic mechanical mode.

\subsection{\label{subsec:dispersiveCoupling} Dispersive coupling}
\begin{figure}[t]
\includegraphics{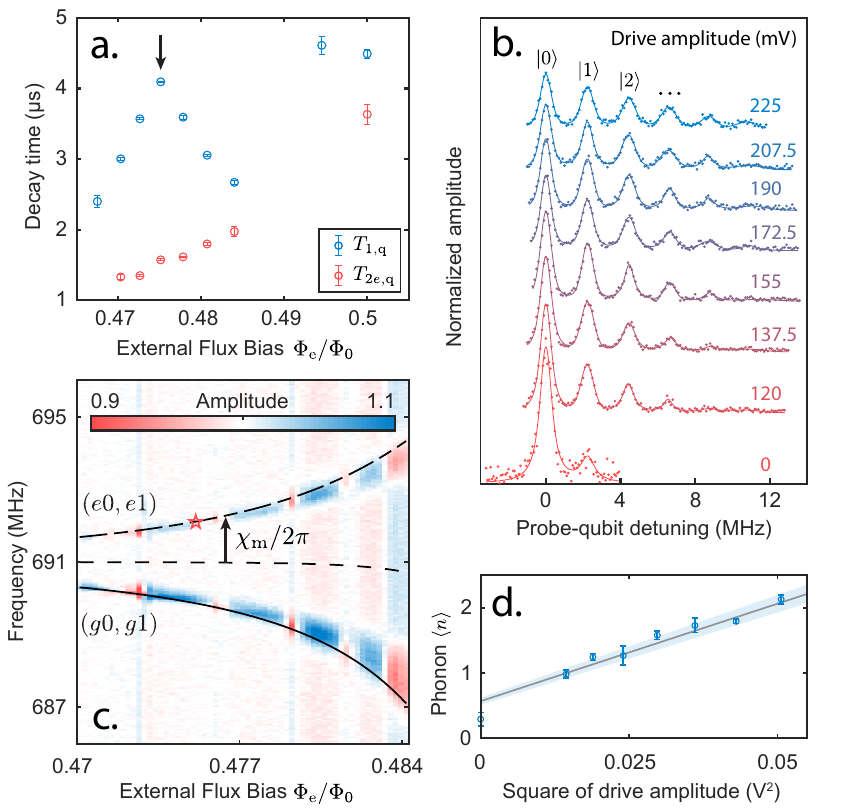}
\caption{\label{fig:f3_phonon_num} \textit{Characterization of phonon number-splitting}. (a) Qubit coherence times as a function of coarsely-stepped flux bias. $\TtwoQe$ is measured using single-pulse echo experiments \cite{Krantz2019} to suppress additional frequency components in the Ramsey signal due to thermal occupation of the mechanics \cite{Wollack2022}. The vertical arrow indicates the bias chosen for number-splitting measurements. (b) Number-splitting spectra for variable coherent drive amplitude. Solid curves show fits to Voigt profiles, and data are rescaled so that the total Fock population is normalized. To compensate for slow frequency drift, spectra are shifted to align centers of the $\ket{0}$ peaks. (c) Spectroscopy of the mechanical mode as the qubit is tuned across the quasi-dispersive regime. For each flux bias, the measured amplitudes are normalized to the mean amplitude over all frequencies, to improve visibility. Upper and lower overlaid curves show calculated transition frequencies continued from Fig. \ref{fig:f2_resonant_coupling}, and their splitting becomes approximately $2\chim$ in the dispersive limit. The pentagram indicates the fitted $\chim$ from (b) relative to average of the outer curves. (d) Calibration of coherent displacement amplitudes extracted from (b). The mean phonon number is calculated from fitted peak areas; further details are given in Appendix \ref{app:phonon_Pn}. The shaded area represents one simultaneous standard error in fit parameters.}
\end{figure}

In the dispersive regime, the qubit frequency is shifted by $2\chim$ for each phonon excitation. We can resolve this splitting in the $(g,e)$ transition by exciting the qubit with pulses at different center frequencies. In our measurement \cite{Arrangoiz-Arriola2019}, we first coherently drive the mechanical resonator to modify the phonon number distribution. We then drive the qubit $(g,e)$ transition while varying the pulse center frequency, after which we measure the qubit state. To select the qubit detuning $\Dcoh$, we step the flux bias from $\Delta/\geg \sim 3$ to $\sim 13$ and measure qubit coherence times, shown in Fig. \ref{fig:f3_phonon_num}a. We choose our qubit-mechanics detuning to be $\Dcoh / \geg\approx 9$ to simultaneously achieve a large detuning (which makes the measurement more QND) and maximize $\ToneQ$. The qubit frequency corresponding to this detuning is $\weg/\tpi \sim 816~\text{MHz}$.
 We then measure the qubit $(g,e)$ spectrum for the case of no driving, and after driving the mechanics with increasing amplitude.
In each case, the mechanics is driven for $T_\text{pump} = 1~\mus$ and the qubit is excited for $T_\text{probe} = 5~\mus$ at each frequency point~\footnote{To measure the qubit spectrum for zero drive amplitude on the mechanics, we drive the qubit for a longer duration $T_\text{probe} = 38~\mus$. This reduces Fourier broadening of the spectral lines, but also reduces the signal-to-noise ratio in the spectrum because the qubit excitation decays as the longer pulse envelope tails off}.
The measured spectra are shown in Fig. \ref{fig:f3_phonon_num}b. In each spectrum we observe multiple peaks, and we observe more peaks at larger drive amplitudes. To estimate the dispersive shift, we calculate the splitting between peaks representing phonon states $\ketm{0}$ and $\ketm{1}$ and obtain $2 \chim / \tpi = 2.23 \pm 0.01~\text{MHz}$. 

The area $A(n)$ under the $n^\mathrm{th}$ peak of the qubit excitation spectrum is proportional to the probability $P(n)$ of phonon state $\ketm{n}$. This allows us to determine the phonon number distribution and to characterize the effect of the coherent driving on the phonon population \cite{Gambetta2006, VonLupke2022}. We fit each spectrum in Fig. \ref{fig:f3_phonon_num}b to a sum of Voigt profiles to obtain the areas $A(n)$, and use the areas to calculate a mean phonon number $\langle n \rangle$ for each drive amplitude. We anticipate that a thermally excited mechanical system with a mean population of $\nbarth$, when coherently displaced by $\alpha$, will result in a mean phonon population of $\langle n \rangle = \nbarth + |\alpha|^2$. Because $\alpha$ should be proportional to drive amplitude, we plot the $\langle n \rangle$ obtained from $P(n)$ as a function of squared drive amplitude, shown in Fig. \ref{fig:f3_phonon_num}d. The linear fit yields a thermal phonon number $\nbarth = 0.57 \pm 0.06$, which for the $690~\text{MHz}$ mechanical mode corresponds to an effective temperature $\Teff = 33 \pm 2~\text{mK}$.

In addition to a shift of the dressed qubit frequency with mechanical excitation number, we expect to see an equivalent shift of the mechanical frequency corresponding to the qubit $(g, e)$ state. Thermal population of $\ketq{e}$ leads to a second peak in the mechanical spectrum. To verify the dispersive model, we measure the mechanical spectrum for varying flux biases with the qubit detuned outside the spectroscopy window. We observe two peaks at frequencies in good agreement with theory, shown in Fig. \ref{fig:f3_phonon_num}c. With the qubit detuned from the mechanics by $\Dcoh$, the dispersive shift $2\chim$ obtained from the qubit peak splitting in Fig. \ref{fig:f3_phonon_num}b also agrees with the mechanical peak splitting. We observe that the mechanical peak splitting decreases faster with increasing $\Delta$ than predicted by the simplified two-level qubit model where $2\chim \approx 2 |\geg|^2/\Delta$. We find (Appendix \ref{app:tuning_fit}) that this behavior is consistent with the contribution of higher qubit levels to the dispersive shift, and similar to what is observed in the transmon-resonator system \cite{Koch2007, Arrangoiz-Arriola2019}. In Fig. \ref{fig:f3_phonon_num}c we also observe regions of decreased peak amplitude. We attribute the reduced signal to resonant couplings between the qubit transition and parasitic mechanical modes at higher frequencies, where the qubit transition frequency is outside of the band gap of the phononic crystal ($595-739~\text{MHz}$). Frequency crowding involving parasitic modes may interfere with control of the target mode and will be addressed in future studies.

\section{\label{sec:phononCoherence} Measuring mechanical lifetimes with frequency modulation}

\begin{figure*}
\includegraphics{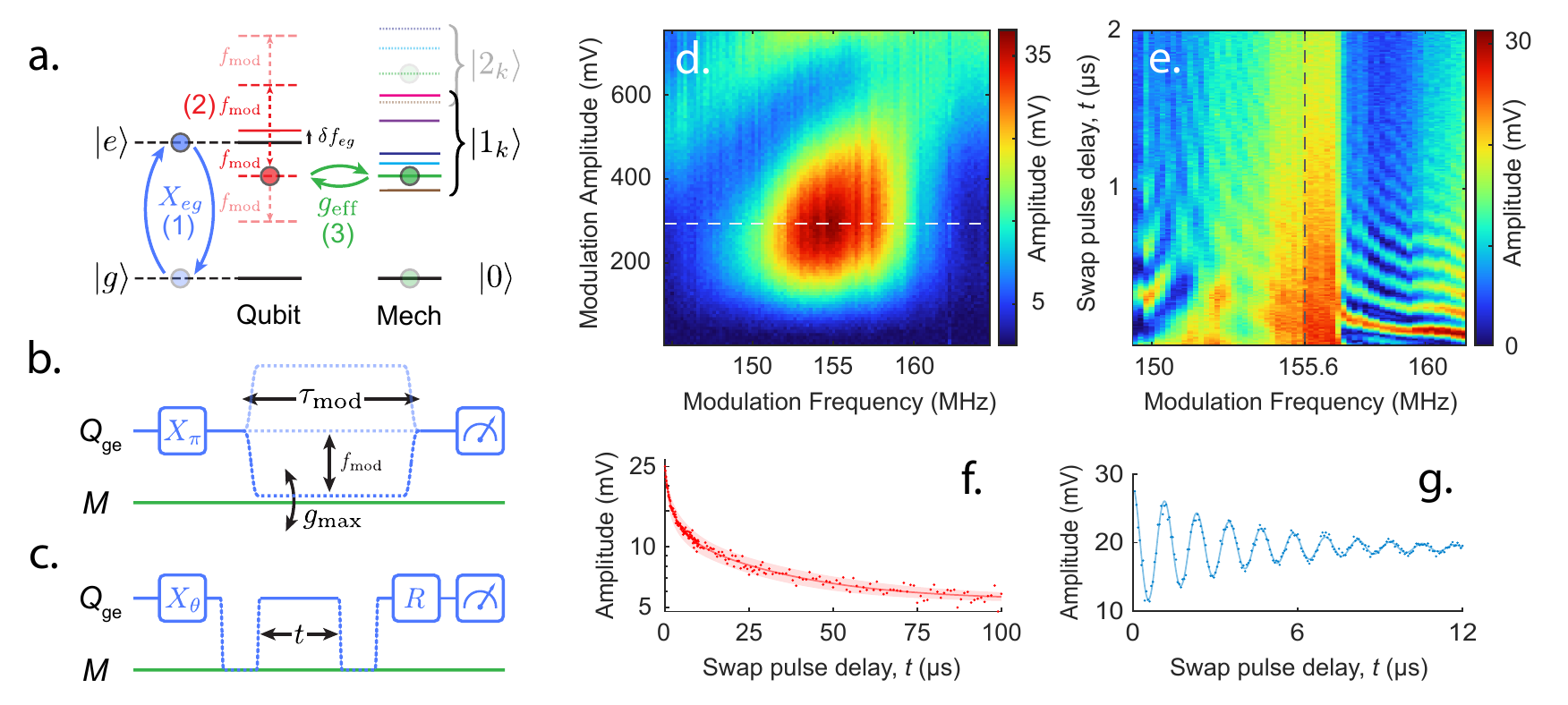}
\caption{\label{fig:f4_freq_mod} \textit{First-order sideband coupling}. (a) Schematic of a Rabi oscillation experiment, driven by flux-modulating the qubit \cite{Naik2017}. Parasitic modes are sketched to emphasize the need for a frequency-selective interaction. Translucent circles represent residual probabilities due to initial thermal populations. (b) Pulse sequence for Rabi experiment. The pulse envelope describes a time-dependent sideband coupling $\geff(\tau)$. (c) Pulse sequence \cite{Wollack2022} for measuring $\ToneM$ or $\TtwoM$. A variable delay time $t$ separates two swap pulses. (d) Qubit response as a function of frequency and amplitude of the flux-modulation pulse in (b). The response is measured relative to a reference experiment where the qubit $\Xsub{\pi}$ pulse is performed with no modulation afterward. The dashed line indicates the modulation amplitude chosen for swap pulses. (e) Qubit response as a function of modulation frequency and short delay times for the pulse sequence in (c), with $R = I$. When the sideband is detuned by $\Delta$ from the target mode, we observe oscillations at frequencies differing from $\Delta$ and attribute them to parasitic couplings \cite{Chu2017}. The dashed line indicates the modulation frequency chosen for swap pulses. (f) $\ToneM$ measurement for the mechanics, using $R = I$. Shading indicates one simultaneous standard error in all fit parameters. (g) $\TtwoM$ measurement for the mechanics, using $R = \Xsub{\pi/2}$.}
\end{figure*}

\subsection{\label{subsec:measSequence} Measurement sequence}

We use the fluxonium to better understand the coherence properties of sub-GHz phononic crystal resonators. We operate the qubit at a large detuning from the mechanical mode and use the measurement sequence shown in Fig. \ref{fig:f4_freq_mod}c. First, we excite the qubit with either a $\pi$ pulse to exchange populations of $\ketq{g}$ and $\ketq{e}$, or a $\pi/2$ pulse to create a superposition. Next we swap the qubit excitation into the mechanical mode by modulating the flux bias at frequencies near the qubit-mechanical detuning, generating an effective coupling rate $\geff$ that depends on the modulation amplitude \cite{Beaudoin2012, Strand2013, Naik2017}. After the swap we allow the system to evolve freely for a time $t$, during which the excitation swapped into the mechanics experiences decoherence from the mechanical environment. We then modulate the flux bias again to swap the excitation back into the qubit. Finally, we measure the qubit state. We measure mechanical energy decay by using a $\pi$ pulse for the initial qubit excitation. To measure mechanical phase decay, we use a $\pi/2$ pulse for the initial qubit excitation, and perform a second $\pi/2$ pulse right before measuring the qubit in the $(g,e)$ basis. For a qubit-mechanical system that begins in the ground state, these experiments measure the coherence properties of a qubit encoded in the $\ketm{0}$, $\ketm{1}$ states of the mechanical oscillator~\cite{Wollack2022}. Due to thermal excitations in our system, the probability of the system beginning in the ground state is reduced, and we expect the experiments to give us some information about the the decay rates of phonon states up to approximately $\ketm{3}$.

We choose the qubit control parameters for mechanical coherence measurements by considering the physical mechanism of the swap operation. Modulating the flux bias at frequency $\wmod$ generates a time-dependent qubit frequency $\weg(t) = \bar{\omega}_\text{eg} + \epsmod \cos(\wmod t + \thetamod)$. The time-dependent frequency creates sidebands of the qubit state $\ketq{e}$, shown in Fig. \ref{fig:f4_freq_mod}a, with frequency spacing equal to $\fmod = \wmod/\tpi$ \cite{Naik2017}. When a sideband is near resonance with the mechanical mode, Rabi oscillations exchange excitations between the qubit and mechanical mode. We couple the first lower sideband to the mechanics by modulating at a frequency $\fmod \sim (\weg - \wmech)/\tpi$, driving Rabi oscillations at a rate $2\geff$ given by,
\begin{equation} \label{eq:mod_geff}
    \geff \approx \geg J_{1}\lp \epsmod/\wmod \rp,
\end{equation}
where $J_{1}$ is the first-order Bessel function of the first kind. To study the target mechanical mode, we avoid unwanted interactions by ensuring that no other qubit sidebands are near resonance with strongly-coupled mechanical modes~\cite{Kervinen2020}. We find that a qubit-mechanical detuning of $\Dswap/\geg \approx 11$ is suitable as it avoids interactions with a second strongly-coupled mode at $950~\text{MHz}$ and reduces the parasitic coupling between the second mode and the first upper sideband.

We calibrate a swap pulse by first exciting the qubit with a $\pi$ pulse, then applying a flux modulation pulse with variable frequency and amplitude (Fig. \ref{fig:f4_freq_mod}b). The modulation pulse duration is fixed at $\taumod = 100~\text{ns}$ including ramps of duration $10~\text{ns}$ on each side. We observe Rabi oscillation as we sweep the amplitude of the pulse as shown in Fig. \ref{fig:f4_freq_mod}d. A large response indicates a significant population transfer from $\ketq{e}$ to $\ketq{g}$, and we choose the modulation amplitude that maximizes the response. An ideal Rabi oscillation pattern is symmetric about the resonant modulation frequency, however we observe a pattern that bends toward higher frequencies as modulation amplitude increases. We attribute this bending to nonlinearity in the flux modulation, causing a small shift in the time-averaged qubit frequency $\bar{\omega}_\text{eg}$ with increasing modulation amplitude \cite{LuThesis2019, Lee2020}. Because this frequency shift is of similar magnitude to the effective coupling $\geff$, we perform a second calibration to verify the resonant modulation frequency. For this calibration we measure energy relaxation of the mechanical mode using the pulse sequence in Fig. \ref{fig:f4_freq_mod}c, following the process described above for delays $t$ up to $2~\mus$, while sweeping the modulation frequency. We observe multiple oscillations in the data (Fig. \ref{fig:f4_freq_mod}e), except at nearly-resonant modulation frequencies $\fmod \sim 155.6 \pm 0.8~\text{MHz}$. 
We choose $\fmod = 155.6~\text{MHz}$ for our swap pulse.

\subsection{\label{subsec:mechCoherence} Mechanical coherence}
After calibrating the swap operation, we measure mechanical coherence using the pulse sequence in Fig. \ref{fig:f4_freq_mod}c while sweeping delays $t$ over a wider range. The result of an energy-relaxation experiment is shown in Fig. \ref{fig:f4_freq_mod}f. We observe a multi-exponential curve that is described well by the sum of three decaying exponentials, similarly to relaxation curves observed for phononic crystal resonators at GHz frequencies \cite{Wollack2022}. We fit a fast decay $\ToneMk{1} = 0.85 \pm 0.13~\mus$, and two slower decays $\ToneMk{2} = 4.11 \pm 0.96~\mus$ and $\ToneMk{3} = 29.6 \pm 4.3~\mus$.
The result of a Ramsey experiment is shown in Fig. \ref{fig:f4_freq_mod}g and is described well by a sinusoid with single-exponential decay. We fit a mechanical dephasing time $\TtwoM = 3.93 \pm 0.17~\mus$. To compare our measured coherence times to recent works in quantum acoustics, we estimate the dispersive cooperativity for amplitude damping \cite{Wollack2022}, $C_{T_1} = (2\chim)^2 \ToneQ \ToneMk{1}$, and for dephasing \cite{VonLupke2022}, $C_{T_2} = (4\chim)^2 \TtwoQ \TtwoM$. With the qubit detuned at $\Dswap$ we perform Ramsey measurements without an echo pulse (Appendix \ref{app:q_dephasing}), fit $\TtwoQ = 0.33 \pm 0.01~\mus$ and $2\chim/\tpi = 1.67 \pm 0.02~\text{MHz}$, and obtain $C_{\Tone, \Ttwo} \approx (330, 570)$. These large cooperativities at large detuning  are competitive with recent works (Table \ref{tab:quac_coop}), despite operating at a larger detuning $\Dswap/\geg \sim 11$, and encourage future studies preparing single-phonon initial states.

\begin{table}
\caption{\label{tab:quac_coop}
\textbf{Dispersive cooperativities in quantum acoustics.} We use the following abbreviations for mechanical resonators: PNC (\textit{phononic crystal}), BAW (\textit{bulk acoustic waves}), SAW (\textit{surface acoustic waves}), and DRUM (\textit{voltage-biased drumhead}). ``+'' represents, ``coupled to''. Cooperativities are rounded to two figures. Values with an asterisk (${}^*$) are predicted using a hypothetical detuning $\Delta$ and device parameters reported in the corresponding reference.
}
\begin{ruledtabular}
\begin{tabular}{lcccr}
\textrm{Experiment} &
\textrm{Year} &
\textrm{$\Delta / \geg$} &
\textrm{$C_{T_1}$} &
\textrm{$C_{T_2}$}\\
\colrule
Fluxonium + PNC [\textit{This Article}] & $2023$ & $11$ & $330$ & $570$\\
Transmon + PNC \cite{Wollack2022} & $2022$ & $-8$ & $490$ & $670$\\
Transmon + BAW \cite{VonLupke2022} & $2022$ & $-7.3$ & $160$ & $590$\\
Transmon + BAW \cite{Kervinen2020} & $2020$ & $-7.3^*$ & $6^{*}$ & $-$\\
Transmon + PNC \cite{Arrangoiz-Arriola2019} & $2019$ & $-6$ & $170$ & $-$\\
Transmon + SAW \cite{Sletten2019} & $2019$ & $11$ & $12$ & $-$\\
Transmon + BAW \cite{Chu2018} & $2018$ & $-7.3^*$ & $160^*$ & $-$\\
Cooper Pair Box + DRUM \cite{Viennot2018} & $2018$ & $172$ & $320$ & $-$\\
\end{tabular}
\end{ruledtabular}
\end{table}

\section{\label{sec:discussion} Conclusions}
We have demonstrated dispersive phonon-resolving measurements of a piezoelectric resonator below $1~\text{GHz}$ using a superconducting qubit in the light-fluxonium regime. We engineered a large qubit-phonon coupling rate within an order of magnitude of the ultra-strong regime \cite{Niemczyk2010} and have leveraged this strong coupling to measure mechanical coherence by flux-modulating the qubit. We observe large dispersive cooperativities of a few hundred (Table \ref{tab:quac_coop}) while operating within the QND regime at a detuning $\Dswap/\geg > 10$. The large cooperativities indicate a strong dispersive interaction between the qubit and mechanics, which exceeds the decoherence rates of both systems. They enable phonon number resolved measurements of our mechanical resonator, and we use these to perform a dissipation and dephasing study of our mechanical system~\cite{Cleland2023}.  Our results open the way for new quantum sensing and information processing schemes with phonons at frequencies below $700~\text{MHz}$. The mechanical frequencies of the resonator in our approach can be readily extended down to $100~\text{MHz}$ by modifying the fluxonium and phononic crystal parameters. Challenges with going to these even lower frequencies include the difficulty in reproducibly fabricating single Josephson junctions with low energies $\EJ/h \lesssim 1.5~\text{GHz}$ \cite{Nguyen2019, NguyenThesis2020}, which limits our ability to operate \textit{light}-fluxonium qubits at arbitrarily low frequencies, and the requirement to etch deeper than $250~\text{nm}$ into lithium niobate to realize thicker phononic crystals at lower frequencies. For the latter, we have recently demonstrated high quality ion-mill etching of lithium niobate with a depth approaching $700~\text{nm}$ for optical devices~\cite{Mishra2022}.
Another experimental limitation of this work was the slow fluctuation in the qubit transition frequency, which we discuss in Appendix \ref{app:drift_tracking}. Frequency fluctuation limited the usable lifetime of calibration measurements to less than one day, preventing us from effectively calibrating a cooling protocol. We suggest experimental modifications to improve frequency stability and implement cooling in Appendix \ref{app:to_improve}. Finally, in contrast to approaches using the transmon, understanding  the phononic crystal response outside of its bandgap is important, particularly to effectively drive fluxonium dynamics beyond $\ketq g$ and $\ketq e$.

\begin{acknowledgments} \label{sec:acknowledgements}
The authors would like to thank K. K. S. Multani, S. Malik, J. F. Herrmann, O. A. Hitchcock, O. T. Celik, M. P. Maksymowych, E. Szakiel, and H. S. Stokowski for useful discussions and assistance during fabrication. The authors would also like to thank and K. Serniak and W. D. Oliver at MIT  Lincoln Laboratory for providing the TWPA. The authors gratefully acknowledge the following sources of financial support for this work: the National Science Foundation CAREER award No.~ECCS-1941826, the U.S. government through the Office of Naval Research (ONR) under grant No. N00014-20-1-2422, the U.S. Air
Force Office of Scientific Research (MURI Grant No. FA9550-
17-1-0002), and funding from Amazon Web Services Inc. E.A.W. acknowledges support from the Department of Defense through the National Defense \& Engineering Graduate Fellowship, while A.Y.C. was supported by the QuaCGR fellowship through the ARO. Part of this work was performed at the Stanford Nano Shared Facilities (SNSF) and at the Stanford Nanofabrication Facility (SNF), supported by the National Science Foundation under award ECCS-2026822.
\end{acknowledgments}

\appendix

\section{\label{app:fabrication} Fabrication}
\begin{figure*}
\includegraphics{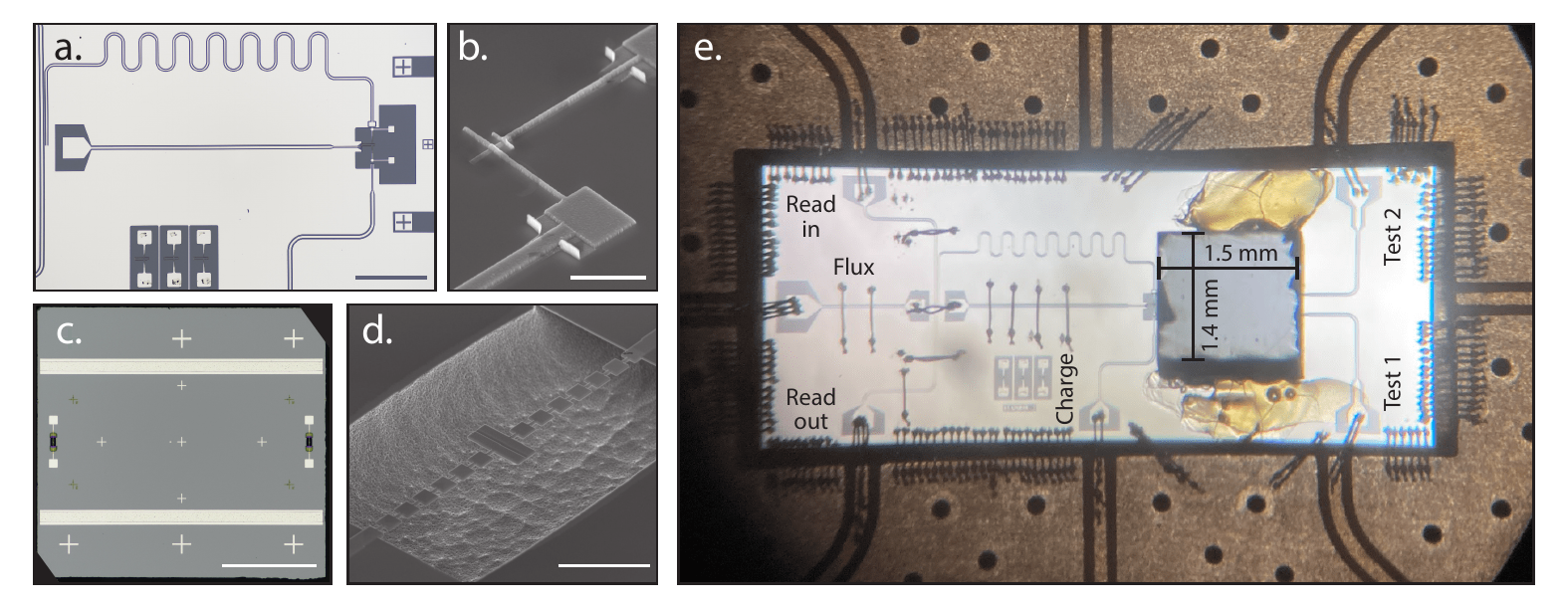}
\caption{\label{fig_app:device_extra} \textit{Extended device images.} (a) Optical micrograph of experimentally active regions on the bottom chip, including the meandered readout resonator. Defects in the aluminum ground plane are associated with debris particles in the photoresist during patterning. (b) Scanning electron micrograph of a representative single Josephson junction with identical geometry to the experimental device. Slight discoloration of the silicon substrate is typical; polymer residue on the aluminum is not ideal. (c) Optical micrograph of the top chip before flip-chip bonding. Corners are truncated by the microscope field of view. No ground plane is used, however a $50~\text{nm}$-thick aluminum film is patterned underneath the $900~\text{nm}$ spacers (long horizontal rectangles) such that the base of the spacers is coplanar with the top surface of the coupling capacitor pads as if a ground plane were present. The top chip is designed with rotational symmetry to enable coupling mechanics on either side to the qubit. (d) Scanning electron micrograph of the representative mechanical resonator from Fig. \ref{fig:f1_device_concept}e, showing the suspended phononic crystal. Scale bars in (a-d) respectively represent (500, 2, 500, 10) $\mum$. (e) Photograph of the experimental device after flip-chip assembly and packaging in a printed circuit board (PCB). Test ports are used to probe a copy of the experimental mechanics. Application of adhesive is intentionally biased toward the test pads to protect the experimental device from unintentional overflow. An example of unintentional overflow can be seen overlapping with the Test $1$ bond pad.}
\end{figure*}
Our device fabrication follows previous methods \cite{Arrangoiz-Arriola2019, Satzinger2018, Wollack2022, Cleland2023}. The mechanical resonators and qubit circuits are fabricated on separate dies and combined in a flip-chip geometry as the final step in fabrication. All electron-beam lithography (EBL) masks are patterned with a JEOL JBX-6300FS ($100~\text{kV}$), and all photolithography masks are patterned with a Heidelberg MLA150 direct-writer ($405~\text{nm}$). All lift-off masks are treated with gentle downstream oxygen plasma to remove polymer residues from interfaces before depositing additional material. An image of the final device is shown in Fig. \ref{fig_app:device_extra}e.

Mechanical oscillators are patterned in thin-film lithium niobate (LN), X-cut with $5~\text{mol}\%$ MgO co-doping, bonded to a silicon $\langle 111 \rangle$ substrate. The fabrication procedure consists of initial film preparation followed by six patterned masks. Starting with an LN thickness of approximately $500~\text{nm}$, samples are thermally annealed for $8$ hours at $500~\text{C}$, then the LN film is thinned to a target of $250 \pm 5~\text{nm}$ by blanket argon ion milling. Mask 1 defines the mechanical structures by EBL using a hydrogen silsesquioxane (HSQ) mask, followed by argon ion milling. Remaining HSQ and redeposited material are removed in a heated bath of dilute hydrofluoric acid followed by baths of piranha and buffered oxide etchant. Mask 2 patterns aluminum electrodes on the LN by EBL and liftoff, and includes the larger coupling pads shown in the inset of Fig. \ref{fig:f1_device_concept}c. Mask 3 patterns aluminum flip-chip alignment marks by photolithography and liftoff. Mask 4 patterns aluminum bandages by EBL and liftoff to ensure galvanic connection of electrodes across the vertical step between silicon and LN. Mask 5 patterns aluminum spacers by EBL and liftoff, with target thickness of $900~\text{nm}$ determining the flip-chip separation distance. Mask 6 performs a masked xenon difluoride dry etch to undercut and suspend the mechanical structures, with mask patterned by EBL.

Qubit circuits are patterned in aluminum on a $525~\mum$ high-resistivity silicon substrate ($\rho > 10~\text{k}\Omega \cdot \text{cm}$). The two-mask fabrication procedure is based on Refs. \cite{KellyThesis2015, Nguyen2019}. Before patterning, the substrate is cleaned in baths of piranha and buffered oxide etchant. Mask 1 patterns qubit electrodes and Josephson junctions by EBL and liftoff, using a Dolan-bridge method \cite{Dolan1977} similar to the patterning of $3\text{D}$ antenna qubits. The geometry of the junction-array inductor is adapted to the asymmetric double-angle evaporation recipe for the T-style single junction \cite{KellyThesis2015}. Mask 2 patterns the ground plane, readout resonator, and all control lines for the qubit, by photolithography and liftoff (150 nm target Al thickness). Our circuit fabrication in this work prioritizes expedience rather than qubit coherence, and we discuss improvements in Appendix \ref{app:to_improve}.

The final fabrication step uses a submicron die bonder (Finetech Fineplacer Lambda) to align the mechanics top-chip to the qubit bottom-chip. The top chip is secured using an adhesive polymer (9:1 ethanol:GE Varnish) applied manually to opposing edges. In our circuit layout, it is necessary to complete the flux control line with a wirebond between on-chip bond pads \cite{Lescanne2020}, which was chosen to simplify routing of coplanar waveguides (CPWs) within the boundaries of the circuit chip ($6.9~\text{mm} \times 2.9~\text{mm}$). The mechanics chip is relatively small ($1.5~\text{mm} \times 1.4~\text{mm}$), to enable manual application of the adhesive without overlapping the superconducting circuits. Manual handling of the mechanics chips after the xenon difluoride etch is minimized, as some previous chips flew away or flipped over before flip-chip bonding due to small agitations on nearby surfaces.

\section{\label{app:exp_setup} Experimental setup}
\begin{figure*}
\includegraphics{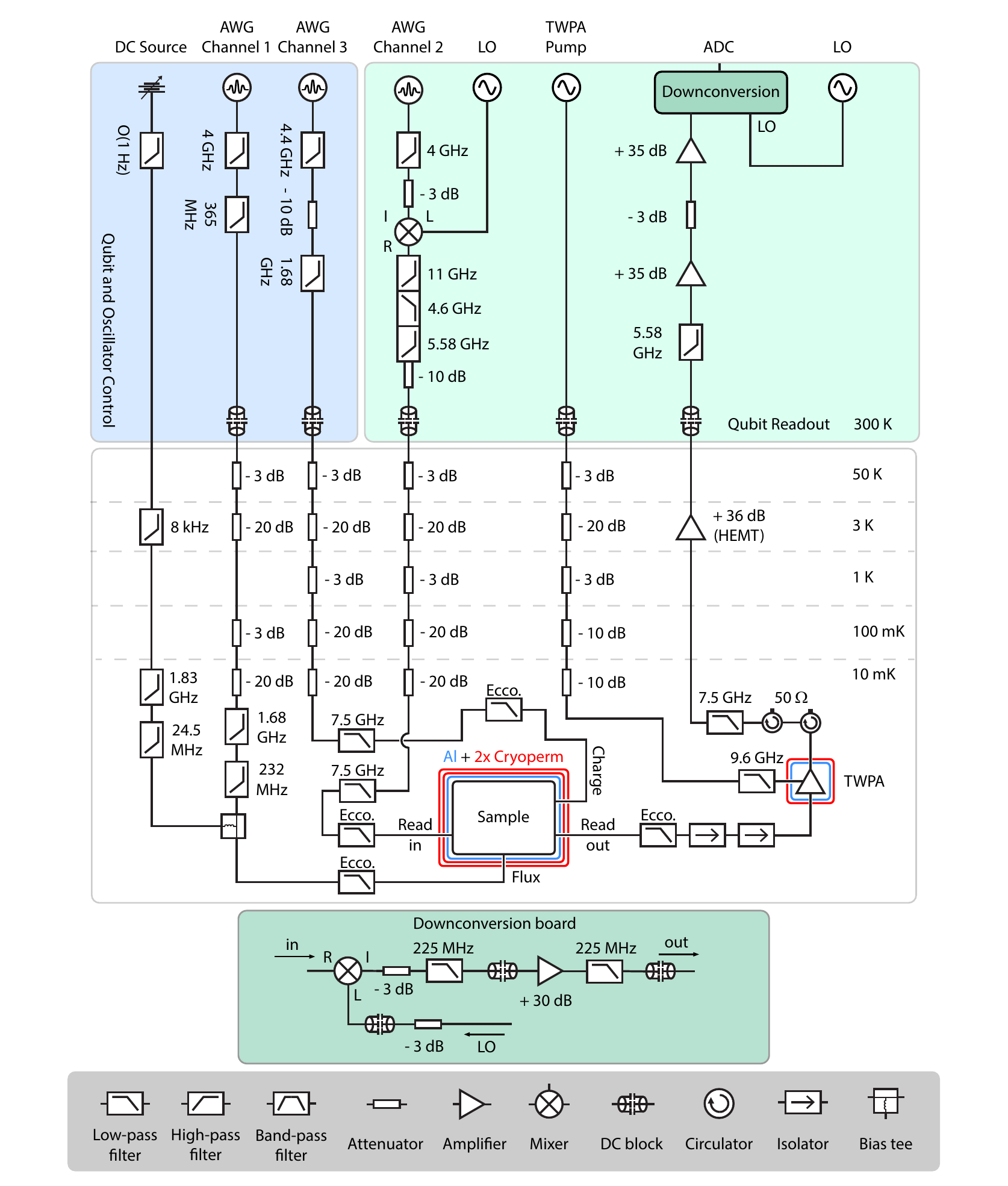}
\caption{\label{fig_app:exp_setup} \textit{Experimental setup.} The sample is located at the mixing-chamber plate of a dilution refrigerator (Bluefors LD250), packaged in a microwave PCB and copper enclosure, and surrounded by cryogenic magnetic shielding. The AWG provides a $10~\text{MHz}$ reference signal to phase-lock all RF instruments, including the ADC. Circulator passbands are $4-8~\text{GHz}$ and isolator passbands are $3-12~\text{GHz}$; both are magnetically shielded with $\mu\text{-metal}$. ``Ecco." denotes coaxial infrared filters made with Eccosorb \cite{FangThesis2015, Krinner2019}, with lowpass cutoffs near $20~\text{GHz}$. The TWPA pump is combined with the readout signal through the $-20~\text{dB}$ port of a directional coupler mounted inside the shielding (RF Lambda RFDC2G8G20, not shown), and the $5.58~\text{GHz}$ lowpass after the HEMT attenuates pump feedthrough to avoid saturating the room-temperature amplifiers with pump power. Notation of this figure follows Refs. \cite{Arrangoiz-Arriola2019, Wang2019}.}
\end{figure*}
The experimental setup is shown in Fig. \ref{fig_app:exp_setup}. We use a $5~\text{GS/s}$ arbitrary waveform generator (AWG) (Tektronix series 5200) for all pulsed experiments in this work. AWG channels (1, 2, 3) respectively output signals for qubit excitation, readout, and flux modulation; channel 4 could be utilized in future work to cool the qubit using few-GHz pulses. Analog upconversion is used to output readout signals near $4.92~\text{GHz}$. All control lines are coaxial except between the DC source at room temperature and the $10~\text{mK}$ plate, which uses a shielded twisted pair with one terminal connected to fridge ground. We use Keysight E8257D sources for local oscillators and to pump a travelling wave parametric amplifier (TWPA). We operate the TWPA \cite{Macklin2015} at a conservative signal-to-noise gain of $15~\text{dB}$ near $4.92~\text{GHz}$ to minimize spurious frequency content, with pump frequency at $6.344~\text{GHz}$. Heterodyne data are collected using analog downconversion of the readout signal to $125~\text{MHz}$, $12-\text{bit}$ digital acquisition at $500~\text{MS/s}$ (AlazarTech ATS9350), and digital down-conversion of one of $\pm 125~\text{MHz}$ to DC.

We attempt to reduce current noise in the flux line using filtering, attenuation, and thermalization of components at the $10~\text{mK}$ stage using copper braid and large-area contact with copper mounts. While the qubit still displays a large pure-dephasing rate (see Appendix \ref{app:q_dephasing}), even a marginal improvement in $\TtwoQ$ is useful for the experiments in this work as the number-splitting measurements would be severely limited by a factor-of-two increase in qubit linewidth. The DC flux line is wired to favor voltage-biasing, in which case the on-chip current noise due to the source is limited by an $8~\text{k}\Omega$ series resistance in the $RC$ filter at the $3~\text{K}$ stage (Aivon Therma-24G). We use an SRS SIM928 for the DC voltage source and add an ultralowpass $RC$ filter across the output \cite{Zhang2021}, contributing another $1~\text{k}\Omega$ of series resistance. Our use of GHz- and MHz-cutoff lowpass filters in the DC flux line and a modified bias tee with capacitor removed from the AC input port also follow Ref. \cite{Zhang2021}. The RF flux line was originally intended for fast DC pulses \cite{Wollack2022}, then reconfigured for RF modulation after frequency drift in similar qubit designs suggested instability in the DC flux bias. 

We use Mini-Circuits components for most of our RF filtering; exceptions include lowpasses at $4.4~\text{GHz}$ (Fairview FMFL-1014), $7.5~\text{GHz}$ (Marki FLP-0750), and $9.6~\text{GHz}$ (Marki FLP-0960). Amplifiers in order from the $3~\text{K}$ stage to downconversion are: Low Noise Factory $\mathrm{LNF-LNC0.3\_14B}$, Miteq AFS3-020018-24-10P, RF-Lambda RLNA05M12GA, and Fairview SLNA-010-30-10-SMA. Isolators and circulators are respectively Quinstar QCI-G0301201AM and QCY-G0400801AM.

\section{\label{app:design} Device design}
\begin{figure*}
\includegraphics{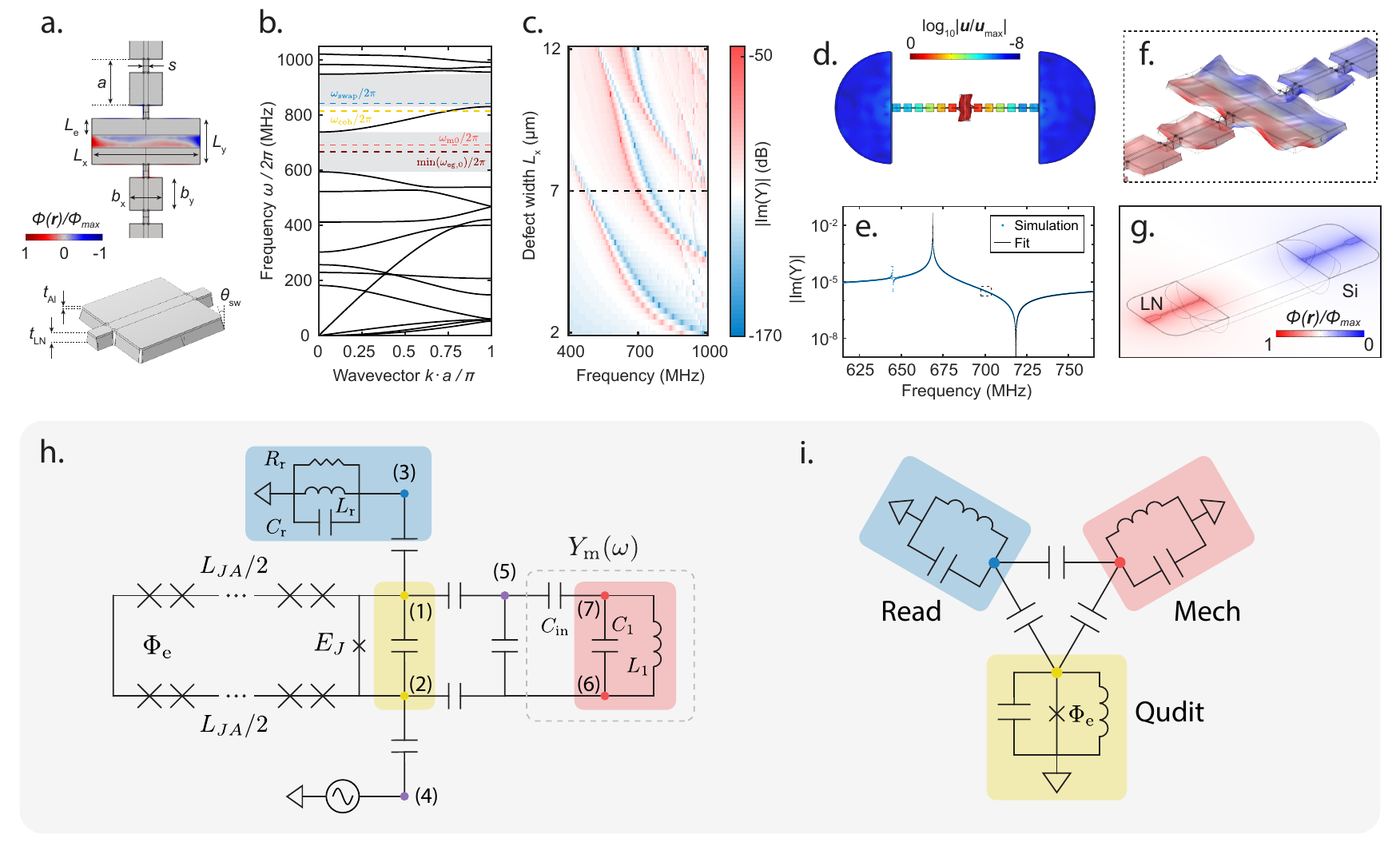}
\caption{\label{fig_app:design} \textit{Design considerations.} (a) Dimensions of phononic crystal cavity, showing electrostatic potential of the target mode from finite-element simulation. Notation follows Ref. \cite{Cleland2023}. (b) Simulated band structure for the phononic crystal mirror cells, with the two largest band gaps shaded. Dashed lines show important qubit and mechanical frequencies in this work, of which only $\wmechzero$ would be targeted at this stage of design. (c) Admittance magnitude across on-defect electrodes as a function of frequency and defect width. To reduce simulation time, the mirror cells are omitted. The dashed line indicates the design width. (d) Target mode simulation of the full resonator, with color indicating the normalized mechanical displacement $\log_{10}|\uvec(\rvec) / \uvec_\text{max}|$. (e) Imaginary admittance magnitude across the electrodes in (d), fit to a single-mode model. The dashed box surrounds a small blip in the admittance associated with a non-design mode near the frequency of the ``parasitic mode" suggested in section \ref{subsec:resonantCoupling} of the main text. (f) Simulated mechanical displacement and electrostatic potential for the parasitic mode indicated in (e). (g) Simulated electrostatic potential of wires on partially-released LN tethers. (h) Nearly-full circuit model used for design. An electrostatic capacitance matrix is simulated for all nodes except $7$ (including capacitances to ground, not shown), then the mechanical admittance model $Y_\text{m}(\omega)$ is inserted using the fit in (e). Shaded boxes identify the three dynamical coordinates in the model. (i) Equivalent circuit (ignoring drives) obtained by reducing the circuit in (h), to be quantized as eqn. \ref{eq_app:H_qmr}.}
\end{figure*}
We model the experimental device with three modes corresponding to qudit, target mechanical mode, and readout mode:
\begin{multline} \label{eq_app:H_qmr}
    \Hhat = 4 \EC \, \nqop^2 - \EJ \cos \lp \phiqop \rp + \half \EL \lp \phiqop + \phie \rp^2 \\
     + \hbar \wmechzero\, \bdagb -i \hbar \gm \nqop \lp \bhat - \bdag \rp \\
     + \hbar \wreadzero\, \rdagr -i \hbar \gr \nqop \lp \rhat - \rdag \rp 
     .
\end{multline}
A typical process for modeling Hamiltonian parameters is outlined in Fig. \ref{fig_app:design}. We design the mechanical resonator is first, as it constrains designs for the coupling circuit.

\subsection{\label{appsub:mech_design} Mechanics}
We design the mechanical resonator using finite-element simulations in COMSOL Multiphysics \cite{Arrangoiz-Arriola2019, Wollack2021, Wollack2022, Cleland2023}. 
The crystal $Z$-axis of the lithium niobate film is oriented perpendicular (horizontal) to the cavity propagation axis (vertical in Fig. \ref{fig_app:design}a). We assume target thicknesses $t_{(\text{LN, Al})} = (250, 50)~\text{nm}$, and we approximate fabrication imperfections using a sidewall angle $\theta_\text{sw} = 12.2^\circ$ and a corner rounding radius of $50~\text{nm}$. To determine the mirror cell dimensions we seek a band gap scaled down in frequency by a factor of $3$ relative to Ref. \cite{Wollack2022}, tripling the lattice constant to $a = 3.0~\mum$. We rescale other planar dimensions by similar factors, yielding a simulated band gap between $595$ and $739~\text{MHz}$. The fractional band gap ($0.216$) is smaller than in previous work, which we attribute to our use of a conservatively large strut width $s = 300~\text{nm}$.

We simulate the electroacoustic admittance across the electrodes to study mechanical resonances. 
To increase the qubit-mechanics coupling strength, we choose the electrode length $L_e = 1.05~\mum$ to be a large fraction of the defect half-length $L_y/2 = 2.95/2~\mum$, and we sweep the defect width $L_x$ over a wide range. To reduce computation time for this sweep, we simulate an isolated defect cell with clamped boundary conditions halfway along the struts leading to the defect, which we find raises predicted resonant frequencies by $10\text{s}$ of MHz. A typical result is shown in Fig. \ref{fig_app:design}c: pairs of admittance poles (red, left) and zeros (blue, right) form curves near the target frequency range, with large pole-zero splitting indicating strong coupling to the electrodes. Three distinct strong-coupling regions are visible in a column around $700~\text{MHz}$. The lowest pole-zero pair corresponds to mode shapes in previous work, resembling half-wavelength shear resonances. The next pole-zero pair corresponds to mode shapes resembling $3/2$ shear wavelengths shown in Fig. \ref{fig_app:design}d for the chosen $L_x = 7.0~\mum$. While we do not observe an increase in the pole-zero splitting using this mode shape, we predict that the increased capacitance between the wide electrodes nevertheless increases the coupling $\gm$ to the qudit charge. After choosing a defect width, we simulate the the full phononic crystal resonator and observe confinement of the mechanical mode displacement to the defect by over $4$ orders of magnitude (Fig. \ref{fig_app:design}d). We fit the admittance near the target mode frequency (Fig. \ref{fig_app:design}e) to an $LC$ model \cite{Wollack2021} and extract circuit parameters shown in Table \ref{tab:design_mech}. The representative mechanical resonator shown in Figs. \ref{fig:f1_device_concept}e and \ref{fig_app:device_extra}d was designed with slightly larger dimensions $L_x = 7.8~\mum$ and $L_e = 1.25~\mum$.

Parasitic mechanical resonances are visible as additional peaks and dips in Fig. \ref{fig_app:design}e. The experimental device was designed by requiring at least $20~\text{MHz}$ of separation between the pole of the target mode and any other extremum in the simulated admittance. While Fig. \ref{fig_app:design}e satisfies this, experiments were still limited in part by frequency-crowding due to non-design modes. An example of a non-design mode is shown in  \ref{fig_app:design}f, corresponding to a barely-visible blip in the admittance at $700.5~\text{MHz}$. This feature was overlooked in designs, where the frequency sweep was not fine enough to detect it. However, we observe a parasitic mode near $697~\text{MHz}$ in experiments (Fig. \ref{fig:f2_resonant_coupling}), and this mode may have interfered with measurements of the qubit-mechanical level structure. Future experiments will benefit from a larger free spectral range between the design mode and other mechanical resonances.

\subsection{\label{appsub:circuit_design} Circuit}
Circuit design amounts to choosing $\EJ$ and $\EL$, and simulating the capacitive network to predict $\EC$. We design the metal geometry for large qubit-mechanics coupling $\geg$ subject to a the following conditions: (1) to observe resonant coupling, the minimum qubit frequency lies below the mechanical frequency, (2) to improve qubit coherence in the dispersive regime, the minimum qubit frequency lies within the primary band gap of the phononic crystal, detuned below the mechanics by several $\geg$, and, (3) to obtain a large readout dispersive shift, the qudit-readout coupling $\gr/\tpi \gtrsim 25~\text{MHz}$ and detuning $|\wreadzero - \omega_{\text{hg}, 0}|/\tpi \sim 100~\text{MHz}$ (in our target regime $|_\text{q} \bra{g} \nqop \ket{h} _\text{q}| \sim 0.3$, so the coupling remains dispersive). The device we implemented experimentally in this work only partially satisfies these conditions. Condition (3) and the second half of (2) are not met. This is due to our use of a top chip with stronger mechanical coupling and smaller capacitive loading relative to designs considered for the bottom chip. Contributions to this effect include removing the top-chip ground plane used in previous works, and decreasing the target flip-chip gap from $1.0~\mum$ to $0.9~\mum$.  We summarize a design process that in principle enables satisfying all the conditions.

The regime of $(\EC, \EJ, \EL)$ targeted in designs follows Ref. \cite{Nguyen2019}, in the neighborhood of qubits (A, D) tabulated therein. A representative example of target parameters is: $\EC/h = 0.7~\text{GHz}$, $\EJ/h = 3.0~\text{GHz}$, $\EL/h = 1.0~\text{GHz}$. The fluxonium regime typically satisfies $1 \lesssim \EJ / \EC \lesssim 10$ and $\EL / \EJ \ll 1$. Here $\EL / \EJ \sim 1/3$ pushes the upper edge of the fluxonium regime such that near half-flux, the harmonic confinement surrounding the double-well potential is steep, and the computational states $\ketq{g, e}$ are not strongly localized in the two wells. This ``light fluxonium" is no longer protected from $\Tone$-type decay as the transition element $ |_\text{q} \bra{g} \nqop \ket{e} _\text{q}| \sim 0.2$ is not strongly suppressed by localization. The low spectrum remains strongly sensitive to $\EJ / \EC$, so we target values of $\EC/h = e^2/2\Csig$ within an accuracy of $\pm 25~\text{MHz}$, corresponding to an accuracy of $\pm 1~\text{fF}$ in the effective qubit capacitance $\Csig$. We therefore attempt to account for all fF-scale contributions to $\Csig$. Starting from a simulated capacitive network sketched in \ref{fig_app:design}h, we consider three additional sources of capacitance. First, an additional electrostatic simulation of the aluminum wires extending across LN tethering structures (\ref{fig_app:design}g) suggests an additional $C_\text{tethers} = 1.0~\text{fF}$, added in parallel to circuit branch $(5, 6)$. Second, for the single junction we assume a plasma frequency $\omega_{J}/\tpi = \sqrt{8 E_{CJ} \EJ}/h \sim 20\text{ to }25~\text{MHz}$, suggesting an additional $C_\text{J} \sim 1 \pm 0.2~\text{fF}$ added across branch $(1, 2)$. Finally, for the array of $N = 74$ junctions we estimate a characteristic impedance $Z_{A} = \sqrt{L_{JA} / C_{A}} \sim 10\text{ to }15~\text{k}\Omega$ using techniques in Refs. \cite{NguyenThesis2020, Viola2015}, adding $C_A \sim 1.2 \pm 0.5~\text{fF}$ across branch $(1, 2)$ \footnote{Theory \cite{You2019} suggests modifying the dependence of the fluxonium Hamiltonian (eqn. \ref{eq:Hq_flxm}) on external flux bias $\phie$ when $\phie(t)$ is time-dependent, as in Appendix \ref{app:rabi_swap} to model frequency modulation. The modification depends on the ratio of capacitances $C_A / C_J$ allocated to the array inductance and single junction. A recent experimental investigation using a fluxonium qubit \cite{Bryon2022} reports results consistent with eqn. \ref{eq:Hq_flxm}, corresponding in theory to the limit $C_A / C_J \ll 1$. In contrast, our estimates of $C_A / C_J$ are of order $1$, however we use these estimates only to predict the charging energy $\EC$, which depends only on $C_A + C_J$. We model the flux modulation as if $C_A / C_J \ll 1$, since we have no direct measurements of $C_A / C_J$.}.

To quantize the circuit, we first reduce the model in Fig. \ref{fig_app:design}h to its three dynamical coordinates using a procedure similar to Ref. \cite{Earnest2018}, yielding the equivalent circuit in Fig. \ref{fig_app:design}i. We omit the readout effective resistance $R_\text{r}$ and replace the voltage driving node $4$ with a short for simplicity in the diagram (extending the calculation to include drives is straightforward). A generic static Lagrangian modeling circuit QED is \cite{Vool2017, Rajabzadeh2022},
\begin{equation} \label{eq_app:L_generic}
    \mathcal{L} = \half \dot{\boldsymbol{\Phi}}^{T} \boldsymbol{C} \dot{\boldsymbol{\Phi}} + U(\boldsymbol{\Phi}; \boldsymbol{\Phie}),
\end{equation}
where $\boldsymbol{\Phi}$ is a vector of node-flux coordinates, $\boldsymbol{\Phie}$ is a vector of external flux biases, $\boldsymbol{C}$ is the Maxwell capacitance matrix, and $U$ is a potential function describing inductors and Josephson junctions. For our relatively simple circuit we identify dynamical coordinates by eye. If the potential function $U$ contains no couplings between a sub-graph $G$ and the rest of the circuit (including ground), there is a conserved charge,
\begin{equation} \label{eq_app:const_Q0}
    \sum_{j \in G} \partial_{\dot{\Phi}_j} \mathcal{L} = \sum_{j \in G} \sum_k C_{jk} \dot{\Phi}_k = Q_{0}.
\end{equation}
If the potential terms within $G$ are of form $U_i(\Phi_a - \Phi_b)$, then defining ${{\Phi}_{ab} \equiv}\,  \Phi_a - \Phi_b$ gives,
\begin{equation} \label{eq_app:const_Q0_2}
    \sum_{j \in G} \lp C_{jb} (\dot{\Phi}_a - \dot{\Phi}_{ab}) + \sum_{k\neq b} C_{jk} \dot{\Phi}_k \rp = Q_{0}.
\end{equation}
Substituting each instance of eqn. \ref{eq_app:const_Q0} or \ref{eq_app:const_Q0_2} into eqn. \ref{eq_app:L_generic} reduces the number of coordinates by one. We use this method to remove node $5$ and substitute ($\Phi_\text{q} \equiv \Phi_1 - \Phi_2$, $\Phi_\text{m} \equiv \Phi_7 - \Phi_6$). Quantization follows from Legendre transform in the reduced coordinates: $H = \sum_{j \in (\text{q, m, r})}  (\partial_{\dot{\Phi}_{j}} \mathcal{L}_\text{red}) \dot{\Phi}_{j} - \mathcal{L}_\text{red}$. A more general method for coordinate transformations in circuit QED is provided in Ref. \cite{Rajabzadeh2022}.

The piezoelectric coupling to the $(g,e)$ transition is given by,
\begin{multline} \label{eq_app:calc_geg}
    \geg = 2 \bqm \sqrt{\wmechzero \EC/\hbar} \, |_\text{q} \bra{g} \nqop \ket{e} _\text{q}| \\
    \leq \half \bqm \sqrt{\wmechzero \, \wegzero},
\end{multline}
where $\bqm = (\Credinv)_\text{qm} / \sqrt{(\Credinv)_\text{qq} (\Credinv)_\text{mm}}$, $\EC = (e^2/2)(\Credinv)_\text{qq}$, and the bound on the charge transition element is derived in Ref. \cite{Pechal2019}. The bound is saturated exactly for linear circuits, while for anharmonic qubits we find transmons achieve $\gtrsim 99\%$ of the bound and light fluxoniums can be engineered to achieve $70-80\%$ of the bound. The utility of strongly-anharmonic circuits for strong coupling is dominated by a large charging energy $\EC$ or equivalently a small capacitance $\Csig = 1/(\Credinv)_\text{qq}$. Using a light fluxonium, we predict an increase in $\bqm$ by a factor of $4\text{ to }5$ relative to a transmon near $700~\text{MHz}$, compensating for the fluxonium's reduced charge element. 

Finally, we consider the largest resonant coupling achievable between a qubit and a piezoelectric mechanical mode. Starting with $\geg/\omega < \bqm/2$, we estimate an upper bound for $\bqm$ using a simplified model where the simulated admittance $Y_\text{m}(\omega)$ in Fig. \ref{fig_app:design}h is shunted by a capacitance $C_\text{q}$ and an arbitrary potential element that sets the qubit on resonance with the mechanics. This describes an ideal coupling circuit where the parasitic capacitance network is eliminated and the qubit is galvanically connected to the mechanical electrodes. In this model, 
\begin{multline} \label{eq_app:beta_qm_ideal}
    \bqm^\text{ideal} = \frac{C_\text{in}}{ \sqrt{(C_\text{q} + C_\text{in})(C_1 + C_\text{in})}} \\
    < \sqrt{\frac{C_\text{in}}{C_1 + C_\text{in}}} = \sqrt{\frac{(8/\pi^2) K^2}{1 - (1-8/\pi^2)K^2}},
\end{multline}
where $K^2$ is the electroacoustic coupling constant \cite{Wollack2021, SarabalisThesis2021}, and the bound is obtained in the limit of negligible $C_\text{q}$. Using this bound we predict that the ultra-strong coupling regime $\geg/\omega > 0.1$ requires $K^2 > 0.05$. For the mechanical admittance in this work we fit $K^2 = 0.16$, and for other defect geometries we fit $K^2 \sim 0.2-0.25$, suggesting that ultra-strong coupling may be possible using an improved coupling circuit. The bound in eqn. \ref{eq_app:beta_qm_ideal} is optimistic, and increasing $\bqm$ in experiments remains a topic of future work.

\begin{table}
\caption{\label{tab:design_mech}
\textbf{Design and test-device parameters for mechanical mode}. Design values for simulation of the target mode shown in Fig. \ref{fig_app:design}d, equivalent circuit parameters, and mode parameters for the test device measured at room temperature (RT) by reflection off test ports ($1, 2$) in Fig. \ref{fig_app:device_extra}, using a VNA and $-45~\text{dBm}$ output power. Table follows Ref. \cite{Cleland2023} and parameter conversions given in Ref. \cite{Wollack2021}.
}
\begin{ruledtabular}
\begin{tabular}{lcr}
\textrm{Description} &
\textrm{Parameter} &
\textrm{Value}\\
\colrule
Phononic crystal pitch & $a$ & $3.0~\mum$\\
Strut width & $s$ & $300~\text{nm}$\\
Mirror cell width & $b_x$ & $2.1~\mum$\\
Mirror cell length & $b_y$ & $2.1~\mum$\\
Defect width & $L_x$ & $7.0~\mum$\\
Defect length & $L_y$ & $2.95~\mum$\\
On-defect electrode length & $L_e$ & $1.05~\mum$\\
LN thickness & $t_\text{LN}$ & $250~\text{nm}$\\
Aluminum thickness & $t_\text{Al}$ & $50~\text{nm}$\\
LN sidewall angle & $\theta_\text{sw}$ & $12.2^\circ$\\
Corner rounding radius &  & $50~\text{nm}$\\
LN mass density & $\rho$ & $4700~\text{kg/m}^{3}$\\
Effective mass & $m_\text{eff}$ & $8.6~\text{pg}$\\
Zero-point displacement & $x_\text{zpf}$ & $1.2~\text{fm}$\\
Zero-point RMS strain & $\bar{\xi}_\text{zpf}$ & $6.3 \times 10^{-12}$\\
LC model coupling capacitance & $C_\text{in}$ & $1.45~\text{fF}$\\
LC model $Y_\text{zero}$ capacitance & $C_{1}$ & $9.42~\text{fF}$\\
LC model $Y_\text{zero}$ inductance & $L_{1}$ & $5.21 ~\mu \text{H}$\\
BVD coupling capacitance & $C_{0}$ & $1.26~\text{fF}$\\
BVD $Y_\text{pole}$ capacitance & $C_\text{m}$ & $0.193~\text{fF}$\\
BVD $Y_\text{pole}$ inductance & $L_\text{m}$ & $293 ~\mu \text{H}$\\
Electroacoustic coupling & $K^2$ & $0.160$\\
Capacitance from LN tethers & $C_\text{tethers}$ & $1.0~\text{fF}$\\
Mode frequency (measured) & $f_{\text{m}0} (RT)/\tpi$ & $678.8~\text{MHz}$\\
Internal quality factor (meas) & $Q_{i}(RT)$ & $995$\\
Coupling quality factor (meas) & $Q_{e1}(RT)$ & $95.9\times10^3$\\
 & $Q_{e2}(RT)$ & $94.3\times10^3$\\
\end{tabular}
\end{ruledtabular}
\end{table}

\section{\label{app:tuning_fit} Fitting tuning spectrum}
\begin{figure}
\includegraphics{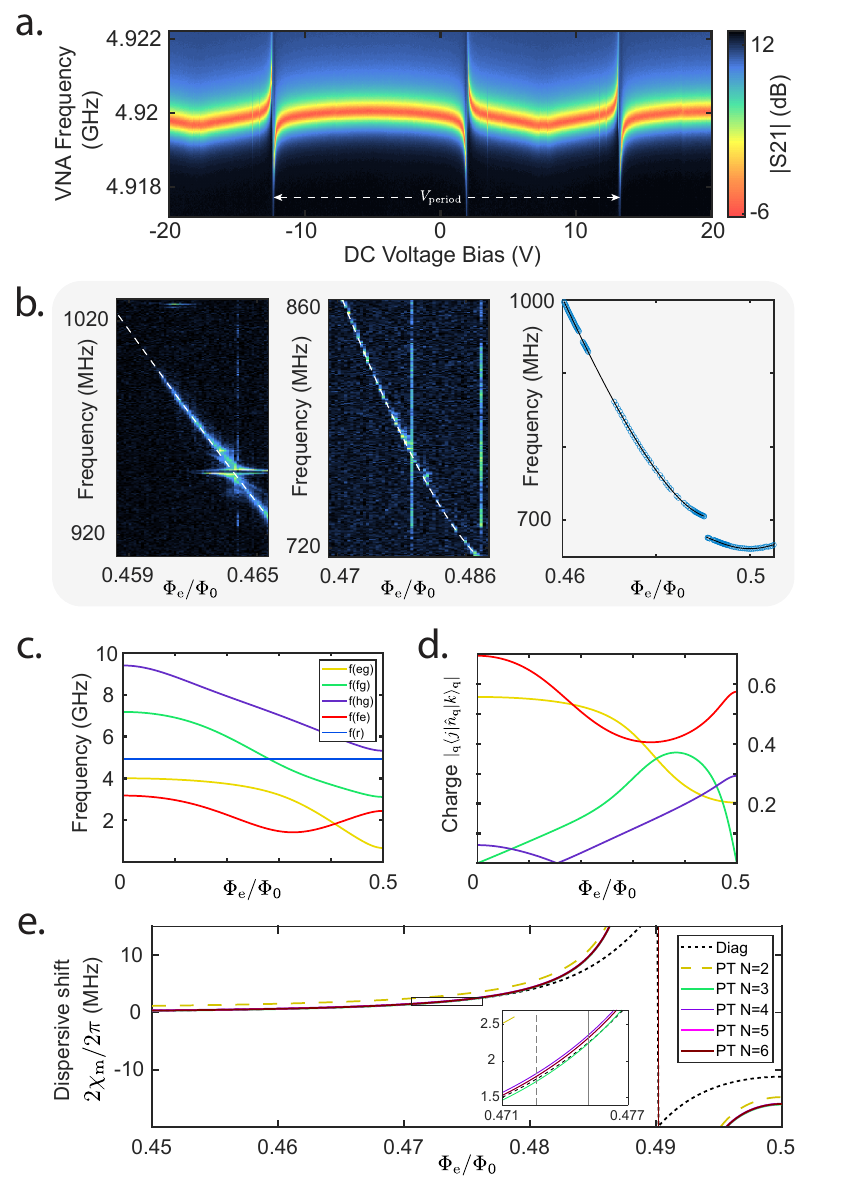}
\caption{\label{fig_app:tuning_fit} \textit{Flux-tuning spectrum}. (a) VNA spectrum of readout mode. The voltage tuning period is estimated from the periodicity of the largest avoided crossing, likely involving the qudit $(g,f)$ transition. (b) Flux-tuning data used to fit qudit energies. Dashed curves are fits overlaid on experimental spectra. The spectroscopy signal disappears abruptly above $1~\text{GHz}$. High-signal vertical bands represent an occasional bug in the measurement chain. (c) Transition frequencies and (d) Charge transition elements for the bare qudit, predicted using fitted energies. The $(e,f)$ transition is included because $\ketq{e}$ has substantial thermal population and because the transition contributes non-negligibly to the qubit-mechanics dispersive shift in eqn. \ref{eq_app:chi_j_partial_PT}. (e) Comparing predicted dispersive shifts using numerical diagonalization (fine-dashed curve) and perturbation theory with varying level truncations. Solid curves represent truncations including $\ketq{f}$ and higher, and agree better with diagonalization compared with truncation at $\ketq{e}$ (coarse-dashed curve). Inset shows the main regime utilized in this work, with (solid, dashed) vertical lines respectively indicating the qubit biases labeled ($\Dcoh$, $\Dswap$) in the main text.}
\end{figure}

Experimental device characterization involves determining parameters in the coupled qudit-mechanics Hamiltonian, given explicitly by,
\begin{multline} \label{eq_app:H_qm}
    \Hhat = 4 \EC \, \nqop^2 - \EJ \cos \lp \phiqop \rp + \half \EL \lp \phiqop + \phie \rp^2 \\
     + \hbar \wmechzero\, \bdagb -i \hbar \gm \nqop \lp \bhat - \bdag \rp ,
\end{multline}
where we have ignored coupling to the readout mode in eqn. \ref{eq_app:H_qmr}. The qudit energies $(\EC, \EJ, \EL)$ are obtained by measuring and fitting the frequency of one of more qudit transitions for variable flux bias $\phie = \tpi \Phie / \Phio$. While for fluxonium it may be preferable to to measure and fit an extended spectrum containing several transitions and/or a large fraction of a flux-tuning period \cite{Earnest2018, Nguyen2019}, we observe inconsistent visibility of GHz-frequency qudit transitions in two-tone spectroscopy and therefore use a restricted fit to the qubit frequency $\weg(\Phie)/\tpi < 1~\text{GHz}$. 

To calibrate external flux, the tuning period with respect to voltage bias is determined independently from measurements of the readout mode with a vector network analyzer (VNA) as the voltage bias is swept (Fig. \ref{fig_app:tuning_fit}a), yielding $V_\text{period} = 25.56 \pm 0.04~\text{V}$. We perform this measurement before all qubit spectroscopy to avoid suspected hysteresis in the qubit frequency associated with larger variations in applied flux. While we observed hysteresis with previous iterations of the device, we do not observe hysteresis for the device in this work. The voltage bias at half-flux is determined as $V_\text{half} = 7.540 \pm 0.01~\text{V}$ from symmetry about the minimum qubit-like frequency in Fig. \ref{fig:f2_resonant_coupling}a.

To fit Hamiltonian parameters given the flux calibration, we first combine measurements of qubit $(g,e)$-like peak frequencies below $1~\text{GHz}$, including the symmetry point at half-flux, and excluding centers of avoided crossings with non-target modes (Fig. \ref{fig_app:tuning_fit})b. Despite the frequency tuning extending far beyond the avoided crossing with the target mode, we include the coupling $\gm$ in the tuning fit because it contributes a large shift to the minimum qubit-like frequency. To expedite fitting we truncate the coupling in eqn. \ref{eq_app:H_qm} to a Jaynes-Cummings model \cite{Girvin2014},
\begin{equation} \label{eq_app:H_int_JC}
    \Hint^\text{JC} / \hbar = \geg \lp \ketq{e} \braq{g} \bhat +  \ketq{g} \braq{e} \bdag\rp,
\end{equation}
where $\geg \equiv \gm |_\text{q} \bra{g} \nqop \ket{e} _\text{q}|$. The approximate qubit-like frequency is,
\begin{equation} \label{eq_app:weg_JC}
    \weg \approx \half \lp \wegzero + \wmechzero + \text{sgn}(\delta_0) \sqrt{\delta_0^2 + 4 \geg^2} \, \rp,
\end{equation}
where $\delta_0 \equiv \wegzero - \wmechzero$ is the bare-basis detuning and $\omega_{\text{eg}, 0}$ is calculated from the first line of eqn. \ref{eq_app:H_qm}. Fitting to eqn. \ref{eq_app:weg_JC} gives: $\EC/h = 0.8016~\text{GHz}$, $\EJ/h = 2.6349~\text{GHz}$, $\EL/h = 0.7966~\text{GHz}$, $\wmechzero/\tpi = 691.71~\text{MHz}$, and $\gm/\tpi = 67.0~\text{MHz}$. We then calculate the eigenfrequencies of the qubit-mechanical system and compare predicted transition frequencies to the spectroscopy measurements in Fig. \ref{fig:f2_resonant_coupling}a. We hold $(\EC, \EJ, \EL)$ fixed to the above values and adjust $\wmechzero$ and $\gm$ to improve agreement between the model curves and data. We evaluate the agreement by eye, so we assume error bars given by ($\wmechzero$) or propagated from ($\gm$) the frequency step in the spectra ($0.25~\text{MHz}$). We find $\wmechzero/\tpi = 691.75~\text{MHz}$ and $\gm/\tpi = 66.6~\text{MHz}$, used for all model calculations. A summary of system parameters and error bars is given in Table \ref{tab:device_params}.

Experimental data suggest that the qubit-mechanics dispersive shift $|2\chim| < 2 |\geg^2 / \Delta|$, which can occur when qudit levels above $\ketq{e}$ contribute to the shift. Second-order perturbation theory gives an expression \cite{Lin2018} for the mechanical frequency shift given qudit state $\ketq{j}$,
\begin{equation} \label{eq_app:chi_j_partial_PT}
    \chi_{\text{m}, j} = \sum_{k \neq j} |g_{jk}|^2 \frac{2 \omega_{kj, 0}}{\ \wmechzero^2 - \omega_{kj, 0}^2},
\end{equation}
where $g_{jk} \equiv i \gm (_\text{q} \bra{j} \nqop \ket{k} _\text{q})$, and $\omega_{kj, 0} \equiv \omega_{k, 0} - \omega_{j, 0}$ are transition frequencies in the bare qubit spectrum (Fig. \ref{fig_app:tuning_fit}c). Eqn. \ref{eq_app:chi_j_partial_PT} includes Bloch-Siegert shifts \cite{Beaudoin2011, Bloch1940}, which are important for the dispersive contribution of qudit transitions $\omega_{kj, 0}$ that are far-detuned from the mechanical frequency $\wmechzero$. The peak separation in number-splitting experiments is approximately $2\chim = \chi_{\text{m}, e} - \chi_{\text{m}, g}$, and the mechanical frequency receives a vacuum shift $\delta \wmech \approx (\chi_{\text{m}, e} + \chi_{\text{m}, g})/2$. In Fig. \ref{fig_app:tuning_fit} we predict $2\chim$ using joint diagonalization and perturbation theory (PT), sweeping the number of states $k$ used in eqn. \ref{eq_app:chi_j_partial_PT}. The PT accuracy improves greatly when the third qudit level $\ketq{f}$ is included, after which including levels above $\ketq{f}$ contributes minimal shift, similarly to a transmon-resonator system \cite{Koch2007}.

\begin{table} 
\caption{\label{tab:device_params}
\textbf{Experimental device parameters}. 
Uncertainties represent one standard error. For qubitlike frequencies $\weg$, the uncertainty describes a typical scale of slow drift in the center of spectroscopy peaks observed over many experiments, with example shown in Fig. \ref{fig_app:drift_comp}. Readout mode parameters were obtained from data shown in Fig. \ref{fig_app:tuning_fit}a.
}
\begin{ruledtabular}
\begin{tabular}{lr}
\textrm{Parameter} &
\textrm{Value}\\
\colrule
$\EC/h$ & $0.8016 \pm 0.0868~\text{GHz}$\\
$\EL/h$ & $0.7966 \pm 0.0380~\text{GHz}$\\
$\EJ/h$ & $2.6349 \pm 0.1334~\text{GHz}$\\
$\wmechzero/\tpi$ & $691.75 \pm 0.25~\text{MHz}$\\
$\gm/\tpi$ & $66.6 \pm 1.2~\text{MHz}$\\
$\geg/\tpi$ (\text{resonant}) & $13.56 \pm 0.25~\text{MHz}$\\
$\wreadzero/\tpi$ & $4.91972 \pm 5\times 10^{-5}~\text{GHz}$\\
$\gr / \tpi$  & $30 \pm 2~\text{MHz}$\\
$\kappa_{\text{r}, e}/\tpi$ & $1.2~\text{MHz}$\\
$\kappa_{\text{r}, i}/\tpi$ & $0.2~\text{MHz}$\\
$\Phie / \Phio (\Dcoh)$ & $0.4751$\\
$\weg/\tpi (\Dcoh)$ & $816 \pm 1~\text{MHz}$\\
$\chim / \tpi  (\Dcoh)$ & $2.23 \pm 0.01~\text{MHz}$\\
$\Phie / \Phio (\Dswap)$ & $0.4726$\\
$\weg/\tpi (\Dswap)$ & $843 \pm 1~\text{MHz}$\\
$\chim / \tpi  (\Dswap)$ & $1.67 \pm 0.02~\text{MHz}$\\
$\ToneQ (\Dswap)$ & $3.57 \pm 0.01~\mus$\\
$\TtwoQ (\Dswap)$ & $0.33 \pm 0.01~\mus$\\
$\TtwoQe (\Dswap)$ & $1.35\pm0.02~\mus$\\
$T_{1\text{m}, (1, 2, 3)}$ & $(0.85, 4.11, 29.6) \pm (0.13, 0.96, 4.3)~\mus$\\
$\ToneMeff$ & $4.79 \pm 0.37~\mus$\\
$\TtwoM$ & $3.93 \pm 0.13~\mus$\\
$\Teff$ & $33 \pm 2~\text{mK}$\\
\end{tabular}
\end{ruledtabular}
\end{table}

\section{\label{app:drift_tracking} Tracking frequency drift}
\begin{figure*}
\includegraphics{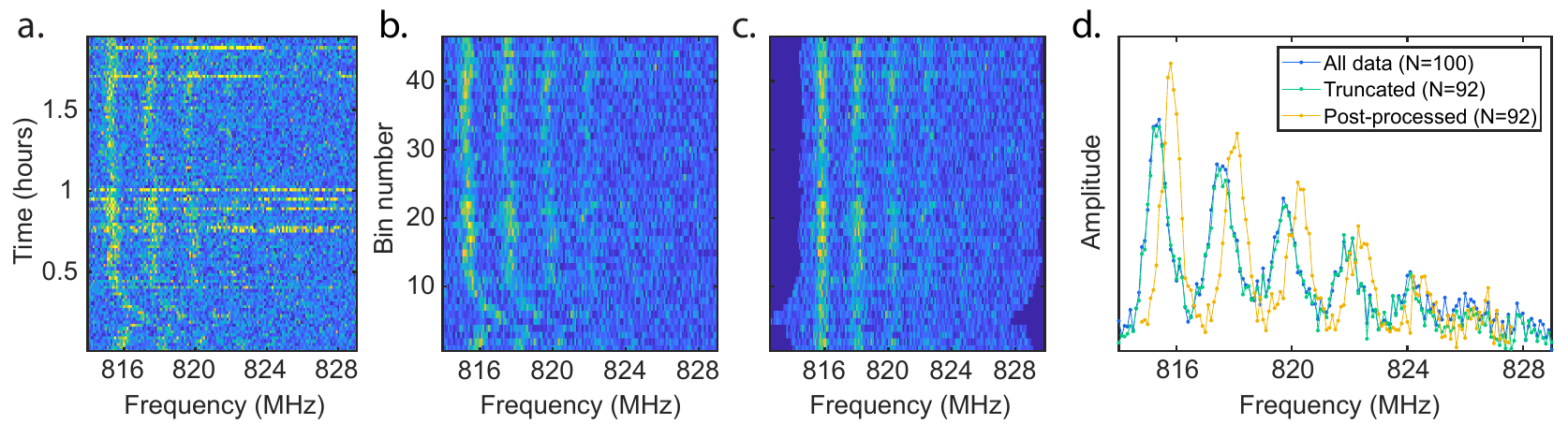}
\caption{\label{fig_app:drift_comp} \textit{Qubit frequency drift.} (a) Raw number-splitting data contributing to Fig. \ref{fig:f3_phonon_num}b for drive amplitude = $225~\text{mV}$ on the mechanical mode. Magnitude of qubit response is shown, with each horizontal slice representing one measurement of the full spectrum averaged over a $70\text{-s}$ interval. $100$ spectra were obtained over nearly $2$ hours, with large drift during the first half hour. The anomalous high-amplitude spectra visible near the center and top of the plot likely represent a bug in the measurement chain, and were observed at rates of $3$ to $7$ per $100$ spectra (b) Truncated and binned data after eliminating $7$ anomalous traces from (a), averaging neighboring spectra in bins of size $2$, and eliminating the remainder. (c) Alignment of binned traces obtained by detecting the highest-amplitude peak, fitting it to a Gaussian profile, and shifting the spectrum by an integer number of the original frequency step. (d) Number-splitting spectra obtained from averaging together the respective spectra in (a, all data), (b, truncated), and (c, post-processed)}
\end{figure*}
When the qubit $(g,e)$ transition is detuned in the dispersive regime above the mechanics, we observe frequency drift on the order of the qubit linewidth over time scales in the tens of minutes. In principle these drifts can be corrected by actively feeding back onto the flux line current to keep the qubit energy fixed. Given that the shifts are small, we find it more convenient to correct this effect in software while post-processing the data. To generate the spectra in Fig. \ref{fig:f3_phonon_num}b of the main text, we partially correct for drift by measuring spectra repeatedly, detecting the frequency of a reference peak in post-processing, and aligning spectra to negate the drift of the reference peak \cite{Arrangoiz-Arriola2019}. An example of this process is shown in Figure \ref{fig_app:drift_comp} for the largest mechanical displacement shown in the main text. We choose the zero-phonon peak of the phonon-number spectra as the reference peak and do not measure additional spectra for peak-tracking purposes. For larger excitation amplitudes the zero-phonon peak is smaller, and we improve the accuracy of peak detection by averaging neighboring spectra together in small bins of $2$ or $3$ before fitting the frequency of the reference peak. This approach benefits from fast repetition of measurements relative to the frequency drift.

Resolution of phonon-number peaks up to $\ketm{4}$ can be seen in Fig. \ref{fig_app:drift_comp}d even without post-processing. The post-processed data improves resolution and symmetry of the peaks and is used for the spectral fits shown in the main text. Using the Hamiltonian parameters extracted in section \ref{app:tuning_fit}, we predict that the dispersive shift $2\chim$ varies by no more than $3\%$ during the observed frequency drifts, contributing a small broadening $\propto n$ to the $\ketm{n}$ peak similarly to phonon loss. We expect this broadening to be negligible relative to the MHz-scale qubit linewidth.

\section{\label{app:q_dephasing} Qubit dephasing}
\begin{figure}
\includegraphics{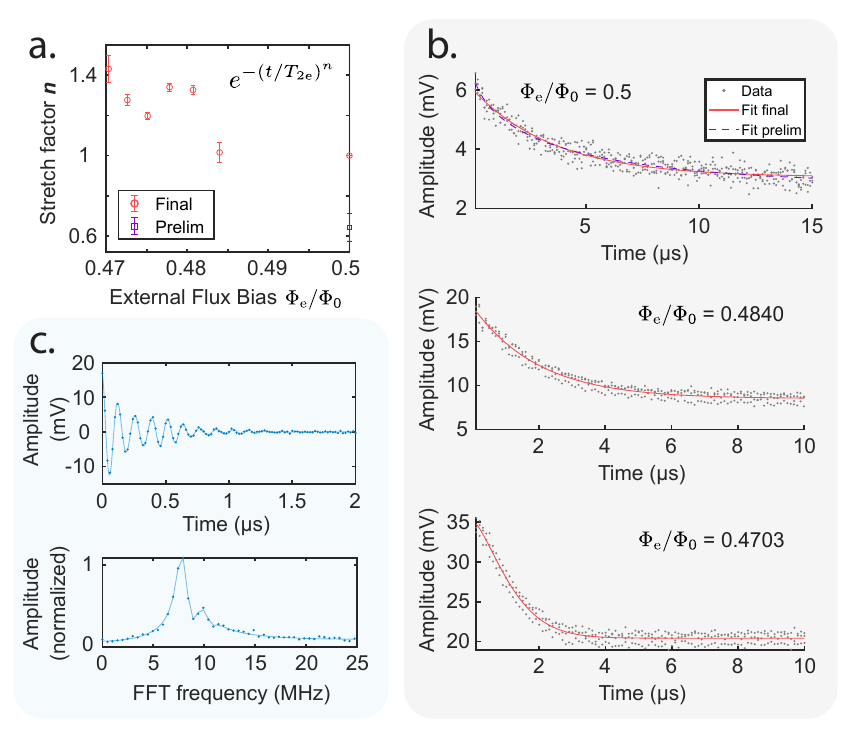}
\caption{\label{fig_app:q_dephasing} \textit{Example echo and Ramsey measurements}. (a) Exponential stretching factors for the $\TtwoQe$ fits shown in Fig. \ref{fig:f3_phonon_num}a of the main text. ``Final" fits were constrained to $n\geq 1$; at half-flux the preliminary unconstrained fit yielded $n \approx 0.64$. (b) Example fits to echo data with different values of $n$. In the upper plot, data and fits at half flux suggest multiexponential decay. (c) Example Ramsey data and fit, including fast Fourier transform (FFT) of each. Weakly-resolved additional peaks suggest thermal phonon populations $P_\text{m}(n = 1, 2)$, motivating our use of echo measurements to suppress dispersive frequency components when the phonon distribution is not of direct interest.}
\end{figure}
Fast qubit dephasing is a major limitation in this work. We fit the maximum $\TtwoQe < 4~\mus$ at half-flux, $\TtwoQe < 2~\mus$ in the dispersive regime, and we observe $\TtwoQe < \ToneQ$ in all cases, suggesting that pure dephasing dominates even with first-order insensitivity to flux noise. We use single-pulse echo measurements for $\TtwoQe$ \cite{Krantz2019} to suppress dispersive frequency components from thermal phonon occupations $P_\text{m}(n = 1, 2) \sim (0.23, 0.09)$, noting that this also refocuses slow dephasing from $1/f$ noise so $\TtwoQe$ tends to exceed the Ramsey $\TtwoQ$. To extract $\TtwoQe$, we fit decay traces to stretched-exponential functions $\exp(-(t/T_{2\text{e}})^n)$ following Ref. \cite{Place2021}, with fitted values of $n$ shown in Fig. \ref{fig_app:q_dephasing}a. Physically, the decay might be described by the product of exponential decay due to white noise and Gaussian decay due to $1/f$-noise \cite{Krantz2019, Zhang2021}, i.e. $\exp(-t / T_C - t^2/T_\phi^2)$. We interpret the stretched-exponential fits as approximations to extract an effective $1/e$ decay time that is easily bounded from the data, and a stretching index $n$ that varies between $1$ for dominant exponential decay and $2$ for dominant Gaussian decay. In principle, $n < 1$ could approximate multi-exponential decay, and a comparison to an exponential decay fit is shown in Fig. \ref{fig_app:q_dephasing}b. 

We also perform non-echo Ramsey measurements with qubit at $\Dswap$ to estimate $\TtwoQ$. Fig. \ref{fig_app:q_dephasing}c shows a time-domain Ramsey fit and its Fourier transform, with model given by \cite{Wollack2022, Cleland2023},
\begin{equation} \label{eq_app:ramsey_dispersive_model}
    S(t) \ = \sum_{n=0}^{N_\text{max}} A_n e^{-t/\TtwoQ}\cos \lb (\omega_0 + 2\chim n)t + \varphi_n \rb,
\end{equation}
where $\lcb A_n, \TtwoQ, \omega_0, 2\chim \rcb$ are fit parameters and we take $N_\text{max} = 2$ after initial fits yielded values of $A_3$ below the noise floor. The phase offset is $\varphi_n = 2\chim n t_\text{d}$, and we set $t_\text{d} = 1.13 \times (\tau_\text{pulse} = 50~\text{ns})$ following simulations in Ref. \cite{Wollack2022} for qubit $X_{\pi/2}$ pulses of the same shape. For the $\TtwoQ$ decay envelope we find more accurate fits using a regular exponential compared to a stretched exponential or Gaussian. Fitting yields $\TtwoQ = 0.33 \pm 0.01~\mus$ and $2\chim/\tpi = 1.67 \pm 0.02~\text{MHz}$. To interpret the short dephasing times, we discuss two contributions to dephasing that may be particularly large for fluxonium qubits and sub-GHz mechanics: strong coupling to a thermal resonator, and $1/f$ flux noise.

\subsection{\label{appsub:dephasing_thermal_mech} Thermal mechanics}
Near half flux $\TtwoQe$ may be limited by phonon number fluctuations from the target mode.  We use the following expression \cite{Clerk2007, Rigetti2012} with caution to estimate the limiting order of magnitude for $\TtwoQe$ due to thermal occupation of the mechanics: $\frac{1}{\TtwoQe} \gtrsim \Gamma_{\phi}^\text{th}$, where,
\begin{equation} \label{eq_app:dephasing_thermal_mech}
    \Gamma_{\phi}^\text{th} \sim \frac{\kapm}{2}\text{Re} \lb \sqrt{\lp1 + i \frac{2\chim}{\kapm} \rp^2 + i\frac{8\chim}{\kapm} \nbarthm} - 1 \rb. 
\end{equation}
Eqn. \ref{appsub:dephasing_thermal_mech} is not limited to $\nbarthm \ll 1$, but we do not anticipate quantitative accuracy because (1) the expression applies to time scales $t \gg (1/\kapm = \ToneM)$ but $\TtwoQe / \ToneM \lesssim 4$, (2) resonator decay is assumed to occur with a single rate $\kapm$ but we observe multiexponential decay, and (3) $|\Delta / \geg| \sim 1.8$ is not in the dispersive regime. Nevertheless, to predict the order of magnitude we use $\nbarthm = 0.57$, $\ToneM = 0.85~\mus$, and $2\chim/\tpi = -11.2~\text{MHz}$ and find $1/\Gamma_{\phi}^\text{th} \sim 1.5~\mus$. For comparison we estimate the qubit pure-dephasing lifetime from measurements, $\TphiQe^{-1} \equiv 1/\TtwoQe - 1/(2\ToneQ)$. This yields $\TphiQe = 6.1~\mus$ at half flux, $4$ times longer than predicted using eqn. \ref{eq_app:dephasing_thermal_mech}. For the dispersive regime accessed in this work, $2\chim / \tpi \geq 1.6~\text{MHz}$, such that eqn. \ref{appsub:dephasing_thermal_mech} predicts $1/\Gamma_{\phi}^\text{th} \sim 1.5-1.6~\mus$ over the entire regime. While the dispersive approximation is more accurate at detunings such as ($\Dcoh$, $\Dswap$) the flux-tuning slope is relatively steep and flux noise becomes a larger contribution to the dephasing rate.

\subsection{\label{appsub:dephasing_1overf} $1/f$ noise}
Our discussion in this section closely follows Refs. \cite{Groszkowski2018, Yan2016, Krantz2019}. Flux-tunable superconducting qubits are broadly affected by $1/f$-type flux noise, with a spectral density of form $\Sphi(\omega) = \Aphi^2 (\tpi \times1~\text{Hz}/|\omega|)^{\gamma_\Phi}$, where $\gamma_\Phi \approx 0.8-1.0$ and $\Aphi^2 \sim (1 \, \mu\Phio)^2/\text{Hz}$. The scaling factor $\Aphi^2$ may be larger if noise from electronics such as the DC bias source is not heavily attenuated, or due to unwanted ground loops. Because qubit coherence is not the main focus in this work, we estimate only the predicted limitation on $\TtwoQ$ and $\TtwoQe$ at the two main static biases used in this work: $(\Dcoh, \Dswap)$. We take $\gamma_\Phi = 1$ for simplicity. The leading-order phase decay in an $N$-pulse Carr-Purcell-Meiboom-Gill (CPMG) experiment is approximately,
\begin{equation} \label{eq_app:dephasing_expr_general_N}
    e^{-\chi_N(t)} = \exp \lb - \frac{t^2}{2} \lp \frac{\partial \weg}{\partial \Phi} \rp^2 \int_{-\infty}^{\infty} g_N(\omega, t) \Sphi(\omega) \frac{d\omega}{\tpi} \rb,
\end{equation}
where the filter functions $g_N$ for experiments in this work are $g_0(\omega, t) = \sinc^2(\omega t/2)$ for Ramsey and $g_1(\omega, t) = \sin^2(\omega t/4)\,\sinc^2(\omega t/4)$ for one echo pulse (approximated as instantaneous). It is typical to exclude frequencies smaller than a cutoff $\omega_c$ if the integral would otherwise diverge. For Ramsey experiments,
\begin{equation} \label{eq_app:dephasing_function_ramsey}
    \chi_{0}(t) \approx t^2 (\Aphi^2 \times\text{Hz}) \lp \frac{\partial \weg}{\partial \Phi} \rp^2 \lp \frac{3}{2} - \gamma + \ln \lp \frac{1}{\omega_c t} \rp \rp, 
\end{equation}
where $\gamma \approx 0.577$ is the Euler constant and we assume $\omega_c t \ll 1$ to ignore terms at $\mathcal{O}((\omega_c t)^2)$; the constant $3/2-\gamma$ can be absorbed as $\ln(2.516/\omega_c t) = \ln(0.400/f_c t)$ in analogy to Ref. \cite{Martinis2003}. The value of $t$ in the logarithm can be set to a representative value on the order of the relevant experimental $\Ttwo$, and the cutoff $\omega_c \sim \tpi/T_\text{exp}$ can be calculated using the total data acquisition time.
For one echo pulse \cite{Ithier2005}, no cutoff is needed at leading order:
\begin{equation} \label{eq_app:dephasing_function_echo}
    \chi_{1}(t) \approx t^2 (\Aphi^2 \times\text{Hz}) \lp \frac{\partial \weg}{\partial \Phi} \rp^2  \ln(2), 
\end{equation}

 We define the pure dephasing time using $-\chi_N(t) \equiv - t^2 / T_{\phi,N}^2$, and for the cutoff logarithm set $t = 5~\mus$ and $\omega_c/\tpi = 1/(600\text{s})$.
 With qubit at ($\Dcoh$, $\Dswap$), $\frac{\partial \weg}{\partial \Phi} \approx \tpi \times (10.67, 11.34) ~\text{GHz} / \Phio$, yielding $T_{\phi,0} = (3.6, 3.4) ~\mus$ and $T_{\phi,1} = (18, 17)~\mus$. The observed $\TphiQe = (1.9, 1.7)~\mus$ are shorter than the calculated $T_{\phi,1}$ by an order of magnitude, and resemble the phonon-fluctuation dephasing time predicted in the previous section \ref{appsub:dephasing_thermal_mech}. However eqn. \ref{eq_app:dephasing_thermal_mech} does not explain the observed trend with tuning away from half flux, where $\TtwoQe$ decreases and the echo decay becomes more Gaussian. Furthermore, the measured ratio between single-echo and Ramsey pure-dephasing times, $\TphiQe / \TphiQ \approx 4.7$, resembles the predicted ratio from eqns. \ref{eq_app:dephasing_function_ramsey} and \ref{eq_app:dephasing_function_echo}: $T_{\phi,1} / T_{\phi,0} \approx 4.9$. These observations could be explained more straightforwardly by a larger noise amplitude $\Aphi^2$ and a smaller phonon-fluctuation dephasing rate. For example, noise amplitudes in the range $\Aphi^2 \sim 1 - 5 ~\mu\Phio^2/\text{Hz}$ have been observed for loops of Josephson junctions \cite{Yan2016, Zhang2021}. Noise amplitudes may increase with increasing geometric aspect ratio $\frac{\text{loop perimeter}}{\text{wire width}}$ \cite{Anton2013}, and in our device this aspect ratio $\sim 200$ is relatively large. Future studies will benefit from quantitatively modeling and reducing pure dephasing.

\section{\label{app:phonon_Pn} Phonon probabilities}
\begin{figure}
\includegraphics{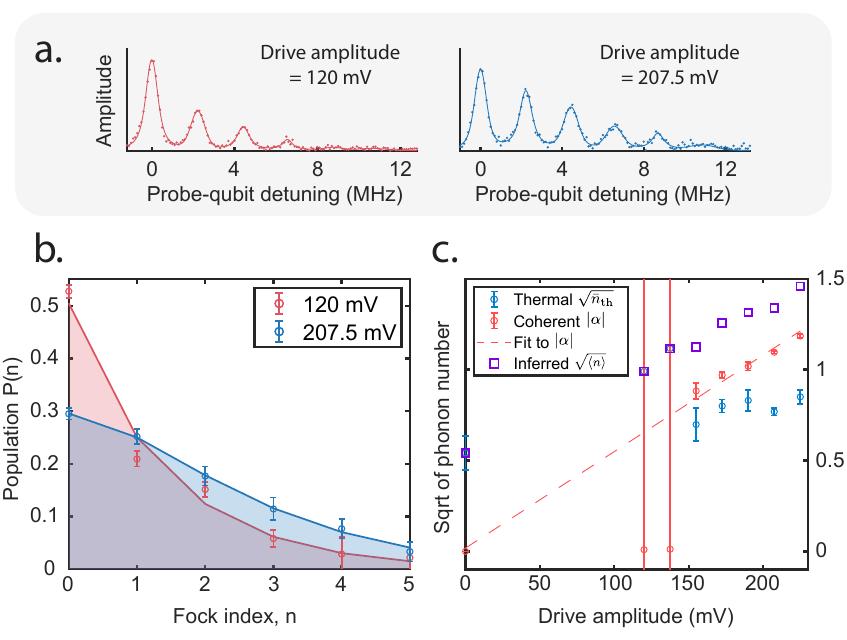}
\caption{\label{fig_app:phonon_num} \textit{Example number-splitting data}. (a) Two example number-splitting traces replotted from Fig. \ref{fig:f3_phonon_num}, representing relatively low and high drive amplitudes. (b) Fock probabilities estimated from data in (a) by fitting relative peak areas to displaced thermal probabilities. Data are normalized such that the fitted distribution would sum to $1$ over all $n$. (c) Results of fits similar to (b) for all drive amplitudes. For the two lowest nonzero amplitudes, the fit did not distinguish an accurate $\alpha$, resulting in very large error bars.}
\end{figure}
Here we describe the processing of phonon number-splitting data shown in Fig \ref{fig:f3_phonon_num} of the main text. Raw spectral data show a raised baseline that increases with mechanics drive amplitude, which could be explained by off-resonant excitation of the qubit or by a small cross-Kerr interaction between the mechanics and readout resonator. To obtain the near-zero baselines shown in Fig. \ref{fig_app:phonon_num}a, we perform reference measurements in which we excite the mechanics with a coherent drive but do not measure the qubit spectrum. We then measure the qubit spectrum following the same coherent drive on the mechanics, and subtract the reference measurement from the measured spectrum.
We fit each spectrum to a model with six Voigt peaks~\cite{VonLupke2022}, each with independent Lorentzian and Gaussian linewidth parameters. We anticipate that the observed lineshapes result from four main broadening mechanisms: white noise due to finite $\ToneQ$ and thermal noise (Lorentzian broadening), $1/f$ flux noise (Gaussian broadening), frequency-shift errors in post-processing (Gaussian by design), and the frequency spectrum of the spectroscopy pulse (sinc-like approximation to Gaussian, for a sinusoidal pulse envelope in time domain). For larger drive amplitudes, we anticipate that a small population $\lesssim 5\%$ in higher phonon levels $ n \geq 6$ is not captured within the spectroscopy window. Because of this, for each drive amplitude we fit the distribution of peak areas to a model, rather than normalizing to the total area of all observed peaks. For the model we choose the Fock distribution of a displaced thermal state \cite{Patel2021},
\begin{multline} \label{eq_app:disp_thermal_Pn}
    P(n) = \bra{n} \hat{D}(\alpha) \hat{\rho}_\text{th} \hat{D}^\dag(\alpha) \ket{n} \\
    = (1-\tau)\tau^n e^{|\alpha|^2 (\tau - 1)} L_{n}\lp-\frac{|\alpha|^2 (\tau-1)^2}{\tau}\rp,
\end{multline}
where $\tau = \exp(-\hbar \wmech / \kB \Teff) = \frac{\nbarth}{\nbarth + 1}$, and $L_n(x)$ are Laguerre polynomials. Two example distributions are compared in Fig. \ref{fig_app:phonon_num}b and a summary for all drive amplitudes is shown in Fig. \ref{fig_app:phonon_num}c. At larger drive amplitudes we observe an expected linear trend for the fitted $\alpha$, though the linear fit (dashed line) would have a larger positive intercept if we did not include zero drive amplitude in the fit. At smaller drive amplitudes the fit does not distinguish a nonzero displacement $\alpha$, so we re-fit to an undisplaced thermal distribution ($\alpha = 0$) and interpret only $\langle n \rangle \approx \nbarth + |\alpha|^2$ quantitatively in the main text. When converted to the same units, the slopes of linear fits in Figs. \ref{fig:f3_phonon_num}d and \ref{fig_app:phonon_num}c agree within one standard error. 

In our number-splitting measurements, the displaced mechanical state decays during the qubit spectroscopy pulse. We use long probe pulses to reduce Fourier broadening, so the bandwidth of the probe pulse resolves the dispersive shift: $2/T_\text{probe} \ll 2\chim/\tpi$. An ideal choice to observe larger phonon distributions would be $T_\text{probe} < \ToneM$, however this requires $\chim \ToneM / \tpi \gg 1$ which is not satisfied in this work. In both this work and Ref. \cite{Arrangoiz-Arriola2019}, $T_\text{probe} > \ToneM$, limiting the size of observed phonon distributions. Despite the mechanical state undergoing significant decay during the measurement, we model the extracted phonon probabilities using displaced thermal states. We motivate this choice by noting that for a resonator undergoing single-phonon loss at rate $\kapm$, a displacement $\alpha(0)$ applied to an initial thermal state decays as $\alpha(t) = \alpha(0) e^{-\kapm t / 2}$ regardless of the thermal occupation \cite{Saito1996}. Measurements at larger drive amplitudes where $|\alpha|^2 > \nbarth$ agree well with this model, as shown in Fig. \ref{fig_app:phonon_num}b for $207.5~\text{mV}$. However for smaller drive amplitudes e.g. $120~\text{mV}$, the model in eqn. \ref{eq_app:disp_thermal_Pn} fits less accurately for $n \in (0, 1, 2)$, passing through none of the error bars. This type of discrepancy appears for the three smallest nonzero drive amplitudes, and was not improved by constraining the fitted displacement $\alpha$ to lie near the linear fit in Fig. \ref{fig_app:phonon_num}c. This behavior suggests a systematic difference between the extracted $P(n)$ and the model for small drive amplitudes, perhaps relating to dephasing in a coupled TLS ensemble \cite{Cleland2023}. More experimental data are needed to evaluate this hypothesis.

\section{\label{app:rabi_swap} Rabi swap}
\begin{figure*}
\includegraphics{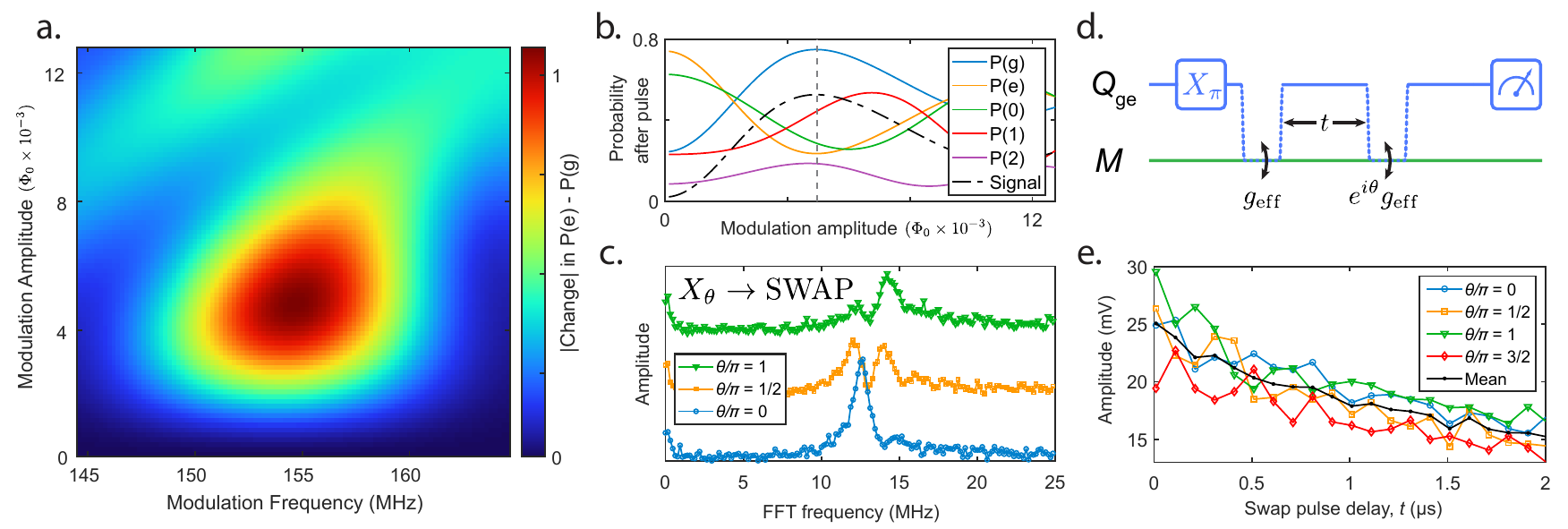}
\caption{\label{fig_app:rabi_swap} \textit{Sideband Rabi swap details.} (a) Simulated ``calibration" for a sideband Rabi swap pulse, modeling experimental data shown in Fig. \ref{fig:f4_freq_mod}d. We plot the magnitude of the change in qubit population asymmetry after the pulse. (b) Simulated qubit and phonon probabilities (unconditioned) as a function of modulation amplitude for fixed frequency $f_\text{mod} = 155.6~\text{MHz}$ used in experiments. ``Signal" represents one half of the value plotted in (a), and for an ideal swap would rise to a maximum of $1$. The vertical dashed line denotes the modulation amplitude chosen for mechanical coherence measurements. (c) FFT of Ramsey data with measurement immediately after the sequence (qubit $X_\theta$, Rabi swap). Lines connecting data points represent only guides to the eye. Peaks representing $\ketm{0,1}$ are visible; we attribute misalignment in frequency between traces to slow flux drift. The signal near DC is spurious, as the mean of each time-domain trace is subtracted before taking the FFT. (d) Pulse sequence for measuring mechanics lifetime $\ToneM$, indicating the modulation phase $\theta$ swept in the $4$-phase average. (e) $\ToneM$ data truncated to short delays $< 2~\mus$ showing few-mV variations in amplitude across the $4$ phases. The mean corresponds to the decay curve fit in Fig. \ref{fig:f4_freq_mod}f of the main text.}
\end{figure*}
We investigate the flux modulation pulse used to swap excitations between the qubit and mechanical mode in section \ref{sec:phononCoherence} of the main text. We perform time-domain simulations using the QuTiP package \cite{Qutip2013} to model an ideal Rabi experiment shown in Fig. \ref{fig:f4_freq_mod}b and compare the result with data shown in Fig. \ref{fig:f4_freq_mod}d. The pulse envelope includes a flat-top of duration $\tau_\text{mod} - 2\tau_\text{r}$ between ramps of duration $\tau_\text{r}$. The upward ramp is given by,
\begin{equation} \label{eq_app:erf_sin_ramp}
    V_\text{up}(t) = V_{0} \left\{\begin{matrix}
 \sin^4 \lp \frac{t}{\tau_d} \rp, & 0 \leq t \leq  \frac{\tau_\text{r}}{2} \\ 
 1 - \sin^4 \lp \frac{\tau_\text{r} - t}{\tau_d} \rp, &  \frac{\tau_\text{r}}{2} \leq t \leq  \tau_\text{r}
\end{matrix}\right. ,
\end{equation}
where $\tau_d = \tau_\text{r}/ \lp 2 \sin^{-1}(2^{-1/4}) \rp$, $(\tau_\text{mod}, \tau_\text{r}) = (100, 10)~\text{ns}$, and the downward ramp follows the upward shape in reverse. We allocate the time-dependent flux to the inductor \cite{You2019, Bryon2022}, such that the drive Hamiltonian is, $\Hhat_d(t) = k V(t) \EL \phiqop$ for some constant $k$. The simulation involves numerically integrating the Lindblad master equation for the qudit-mechanical system,
\begin{equation} \label{eq_app:lindblad_eqn}
    \frac{\mathrm{d} \hat{\rho}}{\mathrm{d} t} = -i \lb \Hhat/\hbar, \hat{\rho} \rb + \sum_{k} \lp \hat{c}_k \hat{\rho} \hat{c}_k^\dag - \half \lcb \hat{c}_k^\dag \hat{c}_k, \hat{\rho}\rcb\rp,
\end{equation}
where $\Hhat = \Hhat_0 + \Hhat_d(t)$, $\Hhat_0$ is given by eqn. \ref{eq_app:H_qm} and we simulate short times up to the pulse duration $\tau_\text{mod}$. We use the following collapse operators: 
\begin{multline} \label{eq_app:c_ops_list}
    \hat{c}_k \in \lp \sqrt{\kappa_{\text{m} \downarrow}}\hat{b}, \, \sqrt{\kappa_{\text{m} \uparrow}}\hat{b}^\dag, \right . \\
    \left . \sqrt{\kappa_{\text{q} \downarrow}}\ketq{g}\braq{e}, \,  \sqrt{\kappa_{\text{q} \uparrow}}\ketq{e}\braq{q}, \, \sqrt{2/T_{\phi, \text{q}}} \ketq{e}\braq{e} \rp, 
\end{multline}
where $\kappa_{\text{m} \downarrow} = (\frac{1 + \nbarthm}{1 + 2 \nbarthm}) \ToneMk{1}^{-1}$, $\kappa_{\text{m} \downarrow} + \kappa_{\text{m} \uparrow} = \ToneMk{1}^{-1}$ and emission/absorption rates for the qubit are assigned similarly. The Hilbert space includes $N_\text{q} = 6$ bare qudit levels obtained by diagonalizing with $N_\text{q, Fock} = 100$, and $N_\text{m} = 10$ bare phonon levels.

Rabi experiments sweeping drive amplitude instead of drive duration often display spurious behavior at large amplitudes, for example due to AC Stark shifts or breakdown of the rotating wave approximation (RWA). Cleaner sinusoidal Rabi chevrons might be observed by sweeping drive duration at a fixed, lower drive amplitude, as performed in many works, for example Refs. \cite{Satzinger2018, Kervinen2020, Wollack2022, VonLupke2022}. We sweep drive amplitude instead of drive duration to circumvent an apparent instrumentation bug where the AWG fails to output flux-modulation pulses longer than $100~\text{ns}$.

Fig. \ref{fig_app:rabi_swap}a shows results of a simulation modeling the Rabi swap calibration attempted in section \ref{sec:phononCoherence}. The initial state is prepared starting with thermal equilibrium at $\Teff = 33~\text{mK}$, followed by an ideal $X_\pi$ pulse modeled as $\hat{U} = \sum_{n}( \ket{e n}\bra{g n} + \text{h.c.})$, where we use $\ket{j n}$ to denote the joint eigenstate with maximum overlap to the bare state $\ketq{j} \otimes \ketm{n}$. Within the dispersive approximation, the readout signal is proportional to the qubit population asymmetry $P(e) - P(g)$, and in experiments we subtract baseline measurements of readout signal obtained by applying the $X_\pi$ pulse without flux modulation afterward. We therefore plot the following quantity,
\begin{equation} \label{eq_app:readout_probability_function}
    \text{Signal} \propto |(P(e) - P(g))_{t = \taumod} - (P(e) - P(g))_{t = 0}|,
\end{equation}
modeling the change in population asymmetry due to the modulation pulse relative to any state preparation done beforehand. In the simulated signal we observe rightward bending similar to the experimental data, and we calibrate the modulation voltage: $k = 1.7\times10^{-5}~\Phio /\text{mVpp}$. The simulated signal does not return to zero between fringes, also consistent with experimental observations. We find that the finite signal between fringes is dominated by dynamics outside the single-excitation subspace $(g1, e0)$ that occur due to thermal excitations. For example, Fig. \ref{fig_app:rabi_swap}b shows unconditioned qubit and phonon probabilities as a function of modulation amplitude using $f_\text{mod} = 155.6~\text{MHz}$. We find that \textit{the maximum readout signal does not coincide with the maximum single-phonon probability} $P(1)$, due to smaller, faster Rabi oscillations occurring in $N$-excitation subspaces that contribute to the unconditioned $P(e)$. The simulated $P(1)$ is approximately $0.45$ at maximum readout signal, while its maximum of $0.54$ occurs at a higher modulation amplitude. Additional work appears necessary to maximize the probability of phonon $\ketm{1}$ without initial cooling. To verify phonon distributions following qubit excitation and a qubit-mechanics swap pulse, we perform Ramsey measurements following Refs. \cite{Wollack2022, Cleland2023}. Fourier transforms of the Ramsey data are shown in \ref{fig_app:rabi_swap}c, and we interpret them qualitatively. We observe two resolved peaks with spacing $2\chim/\tpi \sim 1.6-1.7~\text{MHz}$, and their relative areas appear qualitatively consistent with exchange of population between $\ketm{0}$ and $\ketm{1}$. Data for $\theta/\pi = 1$ appear qualitatively consistent with the prediction $P(1) > P(0) > P(2)$ obtained from Fig. \ref{fig_app:rabi_swap}b at the modulation amplitude where ``Signal'' is maximum.

For $\ToneM$ measurements (Fig. \ref{fig_app:rabi_swap}d), we vary the phase $\theta$ of the second modulation pulse relative to the first. Results of a two-swap experiment can depend on this phase when the target swap interaction is imperfectly calibrated, or when flux modulation drives unwanted interactions. To approximate a DC ringdown curve, we take the  average of the (complex) readout signal over $4$ phases $\theta = (0, \pi/2, \pi, 3\pi/2)$, shown in Fig. \ref{fig_app:rabi_swap}e. This approximately cancels small oscillations between out-of-phase traces for $t \lesssim 1~\mus$.

\section{\label{app:to_improve} Proposed improvements}
Fluxonium qubits often have externally-coupled transitions across a wide range of frequencies. For experiments in this work, the only desired coupling is between the $(g,e)$ transition and the target mechanical mode, however we anticipate that few-GHz transitions such as $(e,f)$, $(g,f)$ and $(g,h)$ also couple to modes in the upper mechanical spectrum, which may complicate readout and cooling protocols. To reduce these spurious couplings we suggest adding a compact, high-impedance lowpass filter between the qubit and mechanics, ideally with a cutoff frequency near $1~\text{GHz}$. Such a filter could be realized with additional piezoelectric design \cite{Cleland2019}, kinetic inductance, or the inductance of a Josephson junction array. To minimize additional processing, filters could be patterned within the LN tethers or by including additional junctions in the qubit metallization.

Fabricating superconducting ground planes and waveguides in the second liftoff mask may decrease $\ToneQ$ and the internal quality factor of the readout mode. In established fabrication procedures \cite{Wollack2022, KellyThesis2015}, ground planes and wiring are typically patterned first on freshly acid-cleaned substrate, followed by Josephson junctions. In superconducting-only systems $\ToneQ$ may also be increased using a sapphire substrate with niobium or tantalum films patterned by etching \cite{Place2021}, and by shortening the perimeter of qubit metal islands \cite{Wang2015}. In our system we observe another, stronger limitation on $\ToneQ$ associated with strong coupling to the target mechanical mode, despite the heterogeneously integrated flip-chip geometry. We measured two additional devices fabricated using the same procedure as this work, but either with reduced qubit-mechanics coupling $\geg/\tpi = 1.3~\text{MHz}$, or without the mechanics top chip. For both devices, we observed an order-of-magnitude longer qubit lifetime $\ToneQ \sim20-100~\mus$. Furthermore, we measured another device composed of a niobium-on-silicon circuit chip and a mechanics top chip, and we observed qubit lifetimes $\ToneQ \sim 2-3~\mus$ similar to this work. While improved circuit fabrication may increase $\ToneQ$, future work will also benefit from understanding limitations on $\ToneQ$ associated with the mechanics chip.

We suggest two modifications to our fabrication of Josephson junctions in this work. First, our patterning of junctions before the ground plane includes baking the junctions at $180^\circ \mathrm{C}$ to prepare the second resist mask, which we find increases the array inductance by up to $30\%$. Although we calibrate the average inductance shift, we observe that the inductance distribution widens and drifts with deviations in bake temperature between fabrication runs. Fabrication control will therefore benefit from avoiding high-temperature bakes after junction fabrication, which may be critical for placing the minimum qubit frequency within the phononic band gap.
Second, our use of the asymmetric T-junction evaporation \cite{KellyThesis2015} for the fluxonium junction array is unusual. The first evaporation at a large angle of $62^\circ$ relative to normal incidence results in larger metal islands between array junctions, increasing the parasitic capacitances associated with these islands. This may lower the frequencies of waveguide-like modes in the junction array \cite{PopThesis2011, Masluk2012, Viola2015}, which contribute to dephasing of the qubit $(g, e)$ transition through a dispersive shift and may couple resonantly to higher qubit transitions. The array island capacitances may be reduced using a symmetric, smaller-angle evaporation \cite{Zhang2021, Place2021} that still yields a small single junction.

Reducing slow drift in the qubit frequency represents a critical improvement for future devices. We anticipate that fabricating superconducting crossovers \cite{Dunsworth2018, Chen2014} as DC shunts across coplanar waveguides may result in improved flux stability, in part by reducing the coupling between loops in the circuit and noisy environmental fields.

High-fidelity quantum operations require cooling the joint qubit-mechanical system. Cooling protocols for superconducting qubits~\cite{Magnard2018, Zhang2021, Somoroff2021} could be extended to cool the mechanics using sideband coupling or fast swaps. We attempted to cool the qubit transition using steady-state driving on the $(e,h)$ qudit transition, relying on emission from the $(g,h)$ transition into the readout mode to cool the qubit. While for previous devices we observed cooling of the $(g,e)$ transition using this approach, we did not observe cooling for the device in this work. We attribute the absence of cooling to a slower Purcell decay of $(g,h)$ through the readout mode, limited by a large detuning $\omega_{hg,0} - \wreadzero$. For future work we consider a more robust cooling protocol \cite{Magnard2018}, in which population of $\ketq{e}$ is transferred to the readout using simultaneous drives on the qudit-readout transitions $(e0,f0)$ and $(f0, g1)$. The method requires calibrating AC Stark shifts on these transitions, due to the large drive amplitudes needed to achieve fast cooling. We anticipate that improved frequency stability will facilitate these calibrations (and therefore cooling) in future studies.

\bibliography{draft_main}

\end{document}

%% file: header_macros.tex
\newcommand*{\half}{\frac{1}{2}}
\newcommand*{\tpi}{2 \pi}

\newcommand*{\lp}{\left (}
\newcommand*{\rp}{\right )}
\newcommand*{\lb}{\left [}
\newcommand*{\rb}{\right ]}

\newcommand*{\lcb}{\left \{}
\newcommand*{\rcb}{\right \}}

\newcommand*{\sinc}{\text{sinc}}

\newcommand*{\kB}{k_{B}}

\newcommand*{\uvec}{\boldsymbol{u}}
\newcommand*{\rvec}{\boldsymbol{r}}

\newcommand*{\Teff}{T_{\text{eff}}}
\newcommand*{\nbarth}{\bar{n}_{\text{th}}}
\newcommand*{\nbarthm}{\bar{n}_{\text{th,m}}}

\newcommand*{\EC}{E_{C}}
\newcommand*{\EL}{E_{L}}
\newcommand*{\EJ}{E_{J}}

\newcommand*{\Csig}{C_{\Sigma}}
\newcommand*{\Credinv}{\boldsymbol{C}^{-1}_\text{red}}

\newcommand*{\wreadzero}{\omega_{\text{r}0}}
\newcommand*{\wmech}{\omega_\text{m}}
\newcommand*{\wmechzero}{\omega_{\text{m}0}}

\newcommand*{\weg}{\omega_\text{eg}} 
\newcommand*{\wegzero}{\omega_{\text{eg}, 0}}

\newcommand*{\chim}{\chi_\text{m}}

\newcommand*{\gm}{g_\text{m}}
\newcommand*{\gr}{g_\text{r}}
\newcommand*{\geg}{g_\text{eg}}

\newcommand*{\bqm}{\beta_\text{qm}}

\newcommand*{\Dcoh}{\Delta_\text{coherent}}
\newcommand*{\Dswap}{\Delta_\text{swap}}

\newcommand*{\fmod}{f_\text{mod}}
\newcommand*{\wmod}{\omega_\text{mod}}
\newcommand*{\epsmod}{\varepsilon_\text{mod}}
\newcommand*{\thetamod}{\theta_\text{mod}}

\newcommand*{\kapm}{\kappa_\text{m}}

\newcommand*{\mus}{\mu\text{s}}
\newcommand*{\mum}{\mu\text{m}}

\newcommand*{\Tone}{T_{1}}
\newcommand*{\ToneQ}{T_{1\text{q}}}
\newcommand*{\ToneM}{T_{1\text{m}}}

\newcommand*{\ToneMeff}{T_{1\text{m}, 1/e}}

\newcommand*{\Ttwo}{T_{2}}

\newcommand*{\TphiQ}{T_{\phi\text{q}}}

\newcommand*{\TtwoQ}{T_{2\text{q}}}
\newcommand*{\TtwoQe}{T_{2\text{e,q}}}
\newcommand*{\TphiQe}{T_{\phi\text{e,q}}}
\newcommand*{\TtwoM}{T_{2\text{m}}}

\newcommand*{\Phie}{\Phi_\text{e}}
\newcommand*{\phie}{\phi_\text{e}}
\newcommand*{\Phio}{\Phi_{0}}

\newcommand*{\Aphi}{A_{\Phi}}
\newcommand*{\Sphi}{S_{\Phi}}

\newcommand*{\geff}{g_\text{eff}}
\newcommand*{\taumod}{\tau_\text{mod}}

\newcommand*{\bhat}{\hat{b}}
\newcommand*{\bdag}{\hat{b}^{\dag}}
\newcommand*{\bdagb}{\bdag \bhat}
\newcommand*{\rhat}{\hat{r}}
\newcommand*{\rdag}{\hat{r}^{\dag}}
\newcommand*{\rdagr}{\rdag \rhat}

\newcommand*{\Heff}{\hat{H}_{\text{eff}}}
\newcommand*{\Hhat}{\hat{H}}
\newcommand*{\Hq}{\hat{H}_\text{q}}

\newcommand*{\Hint}{\hat{H}_\text{int}}

\newcommand*{\sz}{\hat{\sigma}_{z}}

\newcommand*{\nqop}{\hat{n}_\text{q}}
\newcommand*{\phiqop}{\hat{\phi}_\text{q}}

\global\long\def\ket#1{|#1\rangle}
\global\long\def\ketq#1{|#1\rangle_\text{q}}
\global\long\def\ketm#1{|#1\rangle_\text{m}}
\global\long\def\bra#1{\langle#1|}
\global\long\def\braq#1{{}_\text{q}\langle#1|}

\global\long\def\Xsub#1{X_{#1}}
\global\long\def\ToneMk#1{T_{1\text{m}, #1}}